\documentclass[english,journal]{IEEEtran}
\usepackage[T1]{fontenc}
\usepackage[latin9]{inputenc}
\usepackage{color}
\usepackage{float}
\usepackage{mathrsfs}
\usepackage{amsmath}
\usepackage{amsthm}
\usepackage{amssymb}
\usepackage{graphicx}
\usepackage{setspace}

\makeatletter

\providecommand{\tabularnewline}{\\}
\floatstyle{ruled}
\newfloat{algorithm}{tbp}{loa}
\providecommand{\algorithmname}{Algorithm}
\floatname{algorithm}{\protect\algorithmname}

\theoremstyle{plain}
\newtheorem{thm}{\protect\theoremname}
\theoremstyle{definition}
\newtheorem{defn}[thm]{\protect\definitionname}
\theoremstyle{plain}
\newtheorem{lem}[thm]{\protect\lemmaname}
\theoremstyle{remark}
\newtheorem{rem}[thm]{\protect\remarkname}

\usepackage{cite}
\usepackage{times}
\usepackage{algorithmic}

\usepackage{subfig}

\ifdefined\showcaptionsetup
 \PassOptionsToPackage{caption=false}{subfig}
\fi
\usepackage{subfig}
\makeatother

\usepackage{babel}
\providecommand{\definitionname}{Definition}
\providecommand{\lemmaname}{Lemma}
\providecommand{\remarkname}{Remark}
\providecommand{\theoremname}{Theorem}

\begin{document}
\title{Bilinear Subspace Variational Bayesian Inference for Joint Scattering
Environment Sensing and Data Recovery in ISAC Systems}
\author{{\normalsize An Liu, }\IEEEmembership{Senior Member,~IEEE,}{\normalsize{}
Wenkang Xu, Wei Xu, and Giuseppe Caire, }\IEEEmembership{Fellow,~IEEE}{\normalsize\thanks{An Liu, Wenkang Xu, and Wei Xu are with the College of Information
Science and Electronic Engineering, Zhejiang University, Hangzhou
310027, China (email: anliu@zju.edu.cn).

G. Caire is with the Department of Telecommunication Systems, Technical
University of Berlin, 10587 Berlin, Germany (e-mail: caire@tu-berlin.de).}}}
\maketitle
\begin{abstract}
This paper considers a joint scattering environment sensing and data
recovery problem in an uplink integrated sensing and communication
(ISAC) system. To facilitate joint scatterers localization and multi-user
(MU) channel estimation, we introduce a three-dimensional (3D) location-domain
sparse channel model to capture the joint sparsity of the MU channel
(i.e., different user channels share partially overlapped scatterers).
Then the joint problem is formulated as a bilinear structured sparse
recovery problem with a dynamic position grid and imperfect parameters
(such as time offset and user position errors). We propose an expectation
maximization based turbo bilinear subspace variational Bayesian inference
(EM-Turbo-BiSVBI) algorithm to solve the problem effectively, where
the E-step performs Bayesian estimation of the the location-domain
sparse MU channel by exploiting the joint sparsity, and the M-step
refines the dynamic position grid and learns the imperfect factors
via gradient update. Two methods are introduced to greatly reduce
the complexity with almost no sacrifice on the performance and convergence
speed: 1) a subspace constrained bilinear variational Bayesian inference
(VBI) method is proposed to avoid any high-dimensional matrix inverse;
2) the multiple signal classification (MUSIC) and subspace constrained
VBI methods are combined to obtain a coarse estimation result to reduce
the search range. Simulations verify the advantages of the proposed
scheme over baseline schemes.
\end{abstract}


\section{Introduction}

Due to the deployment of wideband orthogonal frequency division multiplexing
(OFDM) waveform and massive multi-input multi-output (MIMO) technologies,
communication signals in future 6G wireless systems tend to have high-resolution
in both time and angular domain. Under such a background, integrated
sensing and communication (ISAC), in which communication signals are
exploited to simultaneously achieve high-speed communication and high-accuracy
sensing, has emerged as a key technology to support many important
applications \cite{LiuFan_survey_signal_process,LiuFan_survey_dual_function,LiuFan_JRC1}. 

There are two basic ISAC scenarios \cite{LiuAn_survey_fundamental_limits}:
broadcast channel (BC) with mono-static sensing and multiple access
channel (MAC) with bi-static sensing. In BC with mono-static sensing,
the base station (BS) acts as both radar transceiver and communication
transmitter. In this case, the BS knows the transmit data and the
sensing problem is relatively easy. In MAC with bi-static sensing,
the BS acts as both radar and communication receivers, while the user
acts as both radar and communication transmitters, as illustrated
in Fig. \ref{fig:Illustration-of-radar}. In this case, the challenge
at BS side is how to perform joint environment sensing, channel estimation,
and data recovery with only limited pilot signals and unknown transmit
data, under possibly imperfect factors such as time offset between
the BS and users, and position errors of users. 

This paper considers the more challenging ISAC scenario of MAC with
bi-static sensing. In particular, we investigate a joint scattering
environment sensing and data recovery problem in the uplink of a massive
MIMO-OFDM ISAC system, where the BS exploits both the uplink pilot
and data signals to jointly detect/localize the scatterers in the
multi-user (MU) propagation environment, estimate the uplink channels,
and recover the data from all users. Some related works are reviewed
below.

\textbf{Joint scattering environment sensing and channel estimation:}
Since multiple users in the same cell share a common scattering environment,
the three-dimensional (3D) scatterers seen by different users have
a partially overlapping structure. The work in \cite{Vincent_joint_sparsity1,LiuAn_joint_sparsity2}
revealed that MU channels exhibit a hidden joint sparsity in the angular-domain
due to the shared scattering environment, which can be exploited to
enhance the channel estimation performance. \cite{LiuFan_JRC1,Huangzhe_JRC2}
concentrated on the partially overlapping structure between radar
targets and communication scatterers, and a joint target sensing and
channel estimation scheme was designed based the joint sparsity of
radar and communication channels. The above works only consider the
oversimplified case when MU or ISAC channels share a partially overlapping
support in the angular domain, while the shared location-domain information
is not fully used. A recent work in \cite{Xu_Turbo-IFVBI} has proposed
a two-dimensional (2D) location-domain sparse channel model to exploit
the partially overlapping support of MU channels for the special case
when all scatterers lies in a 2D area. However, the 3D location-domain
sparse channel modeling is still not fully investigated. Moreover,
the existing works perform joint scattering environment sensing and
channel estimation based on pilot signals only. Nevertheless, data
signals also provides much information about wireless channels that
can help reduce pilot overhead. 

\textbf{Joint channel estimation and data recovery for pure communications:}
In \cite{Cui_JCDE_twostage}, the authors developed a two-stage framework
consisting of a semiblind channel estimator and a semiblind data detector,
where the data detector can use the error statistics to mitigate the
effect of channel estimation error. In \cite{Parker_BiGAMP,Ito_JSCD_BigAMP},
the joint problem is formulated as a bilinear observation model, and
a bilinear generalized approximate message passing (BiGAMP) algorithm
is developed to achieve bilinear sparse signal recovery. A turbo-like
detection approach called GAMP in \cite{Steiner_GAMP} estimated channels
and data consequently and the estimation error of each phase is modeled
as additional additive noise for the following one. In \cite{Thoota_JCDE_VBI1,Thoota_JCDE_VBI2},
a variational Bayesian inference (VBI) algorithm was developed to
obtain the posterior distribution of channels and data alternatively. 

The joint sensing and data recovery problem considered in this paper
is essentially a structured sparse signal recovery problem with a
bilinear observation model and a dynamic grid as well as imperfect
parameters. However, such a problem has not been fully investigated
before, especially for the more practical scenarios of detecting and
localizing the scatterers in a 3D space using both pilot and data
signals, under the consideration of imperfect parameters. Clearly,
the existing algorithms for joint channel estimation and data recovery
cannot be easily extended to solve this problem, and there still lacks
efficient algorithms that have both low complexity and theoretical
performance guarantees. For example, the commonly used BiGAMP algorithm
is heuristic and has no theoretical performance guarantee \cite{Parker_BiGAMP,Ito_JSCD_BigAMP}.
As such, the performance is very sensitive to the properties of the
observation model and fine-tuning of the algorithm parameters. It
is difficult to adjust the algorithm parameters to achieve a stable
performance for the complicated joint sensing and data recovery problem.
The algorithm based on more rigorous theory, such as VBI \cite{Thoota_JCDE_VBI1,Thoota_JCDE_VBI2},
involves high-dimensional matrix inverse. And thus, its complexity
is unacceptable in practice, especially for massive MIMO-OFDM with
huge dimensions of both sparse channels and observations.

In summary, the following challenges have not been fully addressed
in existing works: 1) The modeling and exploitation of the partially
overlapped supports of MU channels (which is referred to as joint
sparsity in this paper) in the 3D location domain remains untapped.
2) For bilinear sparse signal recovery problem with both a dynamic
grid and imperfect factors, there still lacks robust algorithm design
with both theoretical performance guarantees and low complexity. 3)
There still lacks efficient method to handle the ultra-high-dimensional
sparse channel caused by the 3D sparse modeling of massive MIMO-OFDM
MU channels. To overcome those challenges, we propose a joint sensing
and data recovery algorithm based on a 3D location-domain sparse channel
modeling and a novel bilinear subspace variational Bayesian inference
(BiSVBI) algorithm. The main contributions are summarized below.
\begin{itemize}
\item \textbf{3D location-domain sparse channel modeling: }To fully exploit
the partially overlapped structure of 3D scatterers of MU channels,
we propose a 3D location-domain sparse channel representation, where
the entire 3D scattering environment is partitioned into $Q$ dynamic
position grids, and the $q$-th element of the location-domain sparse
channel vector is the channel coefficient corresponding to the scatterer
located at the $q$-th position grid. Then, we propose a three-layer
Bernoulli-Gamma-Gaussian (BGG) prior to capture the joint sparsity
of MU channels in the 3D location domain.
\item \textbf{BiSVBI} \textbf{algorithm:} We model the joint sensing and
data recovery problem as a bilinear structured sparse signal recovery
problem with a dynamic grid and imperfect factors. Then, we propose
an expectation maximization based turbo BiSVBI (EM-Turbo-BiSVBI) algorithm
to solve this problem efficiently, where the E-step performs Bayesian
estimation of the location-domain sparse channels and the M-step refines
the dynamic grid and learns the imperfect factors via gradient update.
In particular, the E-step is based on a novel BiSVBI algorithm by
minimizing the KL-divergence (KLD) between the variational posterior
distribution and the true posterior associated with the bilinear model.
To reduce the complexity, an independent distributed constraint is
imposed on the variational posterior distribution of the sparse channels
to avoid matrix inverse in the calculation of the posterior covariance
matrix. Moreover, a subspace constrained matrix inverse followed by
a gradient update is proposed to replace the high dimensional matrix
inverse in the calculation of the posterior mean. Theoretical analysis
and simulations show that BiSVBI greatly reduces the algorithm complexity
with almost no sacrifice on the performance and convergence speed. 
\item \textbf{Coarse estimation algorithm to further reduce the complexity:}
A coarse estimation algorithm based on spatial multiple signal classification
(SMUSIC) \cite{TST-MUSIC} and subspace constrained VBI (SCVBI) \cite{Liu_SCVBI}
is proposed to narrow down the search range and significantly reduce
the effective dimension of the ultra-high-dimensional sparse signal,
further reducing the complexity of the overall algorithm. 
\end{itemize}

The rest of the paper is organized as follows. In Section \ref{sec:Uplink-System-Model},
we present the system model. In Section \ref{sec:Bilinear-Sparse-Bayesian},
we introduce the 3D location-domain sparse channel model and formulate
the joint sensing and data recovery problem. The proposed EM-Turbo-BiSVBI
algorithm for joint sensing and data recovery and SMUSIC-SCVBI algorithm
for coarse estimation are presented in Section \ref{sec:SEA-Turbo-SBI-Algorithm}
and \ref{sec:SMUSIC-SCVBI-Algorithm-for}, respectively. Finally,
simulation results and the conclusion are given in Section \ref{sec:Applications}
and \ref{sec:Conlusion}, respectively.

\textit{Notation:} Lowercase boldface letters denote vectors and uppercase
boldface letters denote matrices. Let $\left(\cdot\right)^{-1}$,
$\left(\cdot\right)^{T}$, $\left(\cdot\right)^{H}$, $\left\langle \cdot\right\rangle $,
$\textrm{Diag}\left(\cdot\right)$, and $\textrm{Block}\textrm{Diag}\left(\cdot\right)$
represent the inverse, transpose, conjugate transpose, expectation,
diagonalization, and block diagonalization operations, respectively.
$\mathbf{I}_{N}$ is the $N\times N$ dimensional identity matrix.
For a set $\mathbf{\Omega}$ with its cardinal number $\left|\mathbf{\Omega}\right|$,
$\boldsymbol{x}\triangleq\left[x_{n}\right]_{n\in\mathbf{\Omega}}\in\mathbb{C}^{\left|\mathbf{\Omega}\right|\times1}$
is a vector composed of elements indexed by $\mathbf{\Omega}$.

\section{Uplink System Model in Massive MIMO-OFDM ISAC\label{sec:Uplink-System-Model}}

\subsection{System Architecture and Frame Structure}

\begin{figure}[t]
\begin{centering}
\includegraphics[width=80mm]{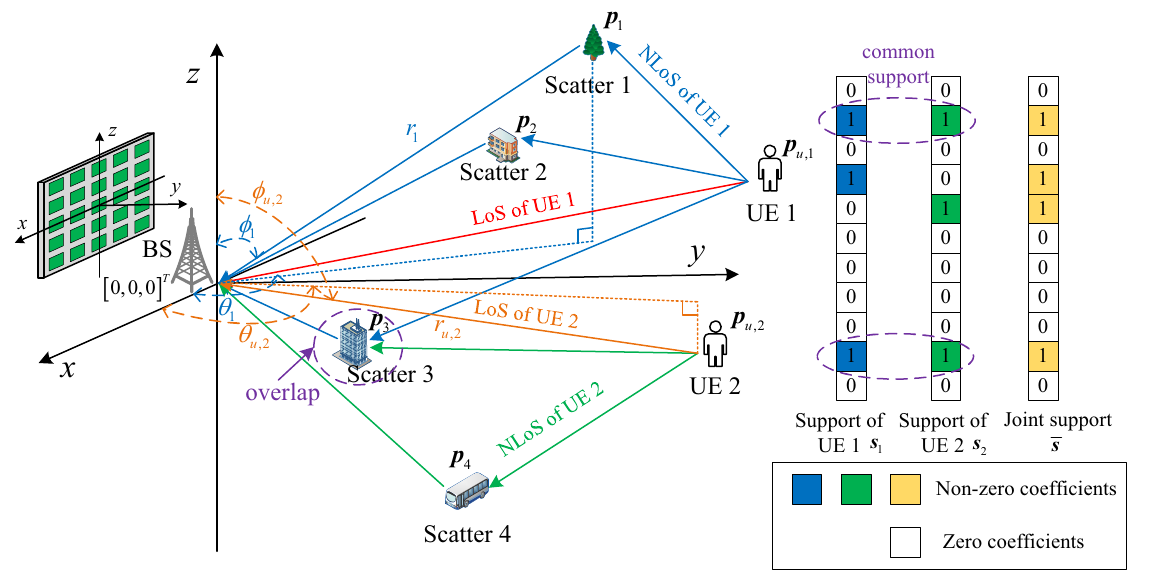}
\par\end{centering}
\caption{\label{fig:Illustration-of-radar}Illustration of the system model,
the location-domain channels, and their non-zero coefficients.}
\end{figure}

Consider the uplink of a massive MIMO-OFDM ISAC system, where one
BS equipped with a uniform planar array (UPA) of $M=M_{x}\times M_{z}$
antennas serves $K$ single-antenna users while sensing the scattering
environment,\footnote{For clarity, we focus on the case when each user has a single-antenna.
However, the proposed algorithm can be readily extended to the case
when each user has multiple antennas and transmits multiple data streams
by treating each antenna/data stream as a virtual user with a single
antenna/data stream.} as illustrated in Fig. \ref{fig:Illustration-of-radar}. In the spherical
coordinate system, the BS\textquoteright s array is located at $\boldsymbol{p}_{b}=\left[0,0,0\right]^{T}$,
and user $k$ is located at $\boldsymbol{p}_{u,k}=\left[\theta_{u,k},\phi_{u,k},r_{u,k}\right]^{T}$,
with $\theta_{u,k}$, $\phi_{u,k}$, and $r_{u,k}$ denoting the azimuth
angle, elevation angle, and range of user $k$, respectively. Suppose
there are a total number of $L$ scatterers indexed by $l\in\mathcal{L}\triangleq\left\{ 1,...,L\right\} $
in the scattering environment, and the $l$-th scatterer is located
at $\boldsymbol{p}_{l}=\left[\theta_{l},\phi_{l},r_{l}\right]^{T},l\in\mathcal{L}$
in a 3-D area $\mathcal{R}$. Besides, we assume there are $L_{k}$
scatterers contributing to the channel paths between user $k$ and
the BS, denoted by the index set $\mathcal{L}_{k}\subseteq\mathcal{L}$.
Moreover, we assume the BS has some prior information about the users\textquoteright{}
location based on existing localization technologies, e.g., Global
Positioning System (GPS) or the previous user localization result.\footnote{The knowledge of the transmitter location (i.e., the user location
in this case) is usually required for performing scatterer localization
\cite{LiuAn_directloc_vehicles,Hong_Scatterer_Loc}.}
\begin{figure}[t]
\begin{centering}
\includegraphics[width=0.8\linewidth]{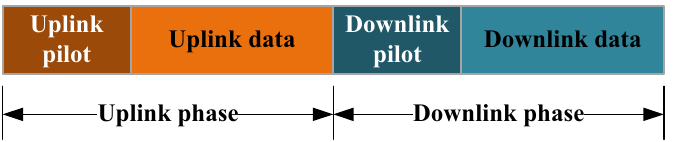}
\par\end{centering}
\caption{\label{fig:Frame-structure}Frame structure of the ISAC system.}
\end{figure}

The time is divided into frames, with each frame containing four phases:
the uplink pilot and data transmission phases, the downlink pilot
and data transmission phases, as illustrated in Fig. \ref{fig:Frame-structure}.
We will focus on the uplink phase that combines the scattering environment
sensing and data recovery into a single procedure. The pilot and data
transmission phases contain $T_{p}$ and $T_{d}$ OFDM symbols indexed
by $t_{p}\in\mathcal{T}_{p}\triangleq\left\{ 1,...,T_{p}\right\} $
and $t_{d}\in\mathcal{T}_{d}\triangleq\left\{ 1,...,T_{d}\right\} $,
respectively, and each OFDM symbol contains $N$ subcarriers indexed
by $n\in\mathcal{N}\triangleq\left\{ 1,...,N\right\} $. In the $t_{p}$-th
pilot symbol, each user $k$ is allocated with a subset of pilot subcarriers
denoted by $\mathcal{N}_{t_{p},k}^{p}\subseteq\mathcal{N}$. In the
$t_{d}$-th data symbol, each user $k$ is allocated with a subset
of data subcarriers denoted by $\mathcal{N}_{t_{d},k}^{d}\subseteq\mathcal{N}$.
The users are allocated with orthogonal pilot subcarriers to avoid
interference. However, multiple users can share the same data subcarrier
to achieve the spatial multiplexing gain.

\subsection{Uplink Signal Model for Joint Sensing and Data Recovery }

The received uplink pilot signal on the $n\textrm{-th}$ subcarrier
of the $t_{p}$-th pilot symbol can be expressed as
\begin{equation}
\boldsymbol{y}_{t_{p},n}^{p}=\boldsymbol{h}_{k_{t_{p},n}^{p},n}u_{t_{p},k_{t_{p},n}^{p},n}+\boldsymbol{z}_{t_{p},n}^{p},
\end{equation}
where $\boldsymbol{h}_{k_{t_{p},n}^{p},n}\in\mathbb{C}^{M\times1}$
denotes the uplink channel vector of user $k_{t_{p},n}^{p}$ on subcarrier
$n$, $u_{t_{p},k_{t_{p},n}^{p},n}$ denotes the pilot signal of user
$k_{t_{p},n}^{p}$, $\boldsymbol{z}_{t_{p},n}^{p}\in\mathbb{C}^{M\times1}$
is the additive white Gaussian noise (AWGN) with variance $1/\gamma$,
and $k_{t_{p},n}^{p}$ is the user allocated with the $n$-th pilot
subcarrier in the $t_{p}$-th pilot symbol.

Similarly, the received uplink data signal on the $n\textrm{-th}$
subcarrier of the $t_{d}$-th data symbol can be expressed as
\begin{equation}
\boldsymbol{y}_{t_{d},n}^{d}=\sum_{k\in\mathcal{K}_{t_{d},n}^{d}}\boldsymbol{h}_{k,n}x_{t_{d},k,n}+\boldsymbol{z}_{t_{d},n}^{d},
\end{equation}
where $x_{t_{p},k,n}\sim\mathcal{CN}\left(x_{t_{p},k,n};0,\sigma_{k,n}^{2}\right)$
denotes the data of user $k$ with $\sigma_{k,n}^{2}$ denoting the
transmit power of user $k$ on subcarrier $n$, $\boldsymbol{z}_{t_{d},n}^{d}\in\mathbb{C}^{M\times1}$
is the AWGN with variance $1/\gamma$, and $\mathcal{K}_{t_{d},n}^{d}$
is the set of users that are allocated with the $n$-th data subcarrier
in the $t_{d}$-th data symbol.

The channel vector $\boldsymbol{h}_{k,n}$ can be modeled as $\boldsymbol{h}_{k,n}=\boldsymbol{h}_{k,n}^{\textrm{L}}+\boldsymbol{h}_{k,n}^{\textrm{NL}}$
with the line-of-sight (LoS) component $\boldsymbol{h}_{k,n}^{\textrm{L}}$
and non-LoS (NLoS) component $\boldsymbol{h}_{k,n}^{\textrm{NL}}$
given by
\begin{subequations}
\begin{align}
\boldsymbol{h}_{k,n}^{\textrm{L}} & =\alpha_{k,0}e^{-j2\pi nf_{0}\left(\tau_{o,k}+\tau_{0}\left(\boldsymbol{p}_{u,k}\right)\right)}\boldsymbol{a}\left(\theta_{u,k},\phi_{u,k}\right),\\
\boldsymbol{h}_{k,n}^{\textrm{NL}} & =\sum_{l\in\mathcal{L}_{k}}\alpha_{k,l}e^{-j2\pi nf_{0}\left(\tau_{o,k}+\tau\left(\boldsymbol{p}_{l},\boldsymbol{p}_{u,k}\right)\right)}\boldsymbol{a}\left(\theta_{l},\phi_{l}\right),
\end{align}
\end{subequations}
where $\alpha_{k,0}$ and $\alpha_{k,l}$ are the complex gain of
the LoS channel path and the $l$-th NLoS channel path, respectively,
which are treated as random variables to facilitate Bayesian algorithm
design. $f_{0}$ is the subcarrier spacing, $\tau_{o,k}$ denotes
the timing synchronization error between user $k$ and the BS, $\tau_{0}\left(\boldsymbol{p}_{u,k}\right)=\frac{r_{l}}{c}$
is the propagation delay of the LoS path from user $k$ to the BS
with $c$ denoting the speed of light, 
\begin{align}
\tau\left(\boldsymbol{p}_{l},\boldsymbol{p}_{u,k}\right) & =\frac{\sqrt{r_{u,k}^{2}+r_{l}^{2}-2r_{u,k}r_{l}g\left(\boldsymbol{p}_{l},\boldsymbol{p}_{u,k}\right)}}{c}+\frac{r_{l}}{c},\\
g\left(\boldsymbol{p}_{l},\boldsymbol{p}_{u,k}\right) & =\sin\left(\phi_{u,k}\right)\sin\left(\phi_{l}\right)\cos\left(\theta_{u,k}-\theta_{l}\right)\nonumber \\
 & +\cos\left(\phi_{u,k}\right)\cos\left(\phi_{l}\right),
\end{align}
 is the propagation delay of the NLoS path from user $k$ to the BS
via the $l$-th scatterer, and $\boldsymbol{a}\left(\theta,\phi\right)$
is the array response vector for the UPA. Note that any multiple-bounce
NLoS path of user $k$ with arbitrary angles $\left(\theta,\phi\right)$
and delay $\tau$ can be equivalent to a single-bounce NLoS path corresponding
to a virtual scatterer at $\boldsymbol{p}=\left(\theta,\phi,r\right)$,
where $r$ is the solution of 
\begin{equation}
\sqrt{r_{u,k}^{2}+r^{2}-2r_{u,k}rg\left(\boldsymbol{p},\boldsymbol{p}_{u,k}\right)}+r-c\tau=0.
\end{equation}

In the uplink of the massive MIMO-OFDM ISAC system, both the received
pilot and data signals $\boldsymbol{y}_{t_{p},n}$ and $\boldsymbol{y}_{t_{d},n}$
are exploited to sense the scattering environment, i.e., detect the
number of active scatterers in $\mathcal{L}_{k},\forall k$ associated
with each user location $\boldsymbol{p}_{k},\forall k$, and estimate
the locations of all scatterers $\boldsymbol{p}_{l},l\in\mathcal{L}$.
In addition, the time offsets $\tau_{o,k}$'s and the user locations
$\boldsymbol{p}_{u,k}$'s will also be estimated/refined to mitigate
those imperfect factors. The sensing results can be used to reconstruct
the channel vectors for data recovery and the recovered data can be
used to further enhance the sensing performance. As such, sensing
and communications can enhance each other by performing a joint sensing
and data recovery algorithm at the BS.

\section{Bilinear Sparse Bayesian Inference Formulation\label{sec:Bilinear-Sparse-Bayesian}}

In this section, we first obtain a location-domain sparse representation
of the channel vectors. Then, we introduce a three-layer BGG prior
to capture the joint sparsity of MU channels. Finally, we formulate
joint sensing and data recovery as a bilinear sparse Bayesian inference
problem.

\subsection{3D Location-Domain Sparse Representation}

We introduce a grid-based method to obtain a sparse representation
of the NLoS channel. Specifically, let
\begin{equation}
\left(\bar{\boldsymbol{\theta}},\bar{\boldsymbol{\phi}}\right)=\left\{ \left(\bar{\theta}_{i,j},\bar{\phi}_{i,j}\right)\mid i=1,...,Q_{1},j=1,...,Q_{2}\right\} ,\label{eq:uniangrid}
\end{equation}
denote $Q_{a}\triangleq Q_{1}\times Q_{2}$ angle grid points and
define $g_{1}\left(\theta,\phi\right)=\cos\left(\theta\right)\sin\left(\phi\right)$,
$g_{2}\left(\phi\right)=\cos\left(\phi\right)$. Then, the angle grid
points are chosen to satisfy
\begin{align}
g_{1}\left(\bar{\theta}_{i,j},\bar{\phi}_{i,j}\right) & =-\frac{Q_{1}-1}{Q_{1}}+\frac{2(i-1)}{Q_{1}},\nonumber \\
g_{2}\left(\bar{\phi}_{i,j}\right) & =-\frac{Q_{2}-1}{Q_{2}}+\frac{2(j-1)}{Q_{2}},\label{eq:defg}
\end{align}
for $i=1,...,Q_{1}$ and $j=1,...,Q_{2}$. In other words, the angle
grid is uniform in the transformed angular domain $g_{1}\left(\theta,\phi\right),g_{2}\left(\theta\right)$.
Besides, the $Q_{r}$ range grid points $\left\{ \bar{r}_{1},\bar{r}_{2},...,\bar{r}_{Q_{r}}\right\} $
are uniformly placed in the range domain itself. Finally, the 3-D
uniform position grid $\left\{ \overline{\boldsymbol{r}}_{1},\ldots,\overline{\boldsymbol{r}}_{Q}\right\} \subset\mathcal{R}$
with $Q\triangleq Q_{a}\times Q_{r}\gg L$ positions is given by the
$Q_{a}\times Q_{r}$ combinations between the $Q_{a}$ angle grid
points and $Q_{r}$ range grid points.

In practice, the true positions of scatterers usually do not lie exactly
on the grid. Therefore, it is essential to overcome the position mismatches
for high-resolution localization. One common solution is to introduce
a dynamic position grid, denoted by $\boldsymbol{r}\triangleq\left[\boldsymbol{r}_{1};\ldots;\boldsymbol{r}_{Q}\right]$
with $\boldsymbol{r}_{q}=\left[\theta_{q},\phi_{q},r_{q}\right]^{T}$,
instead of only using a fixed position grid. In this case, there always
exists a grid $\boldsymbol{r}^{*}$ that covers the true positions
of all scatterers. In general, the uniform grid is chosen as the initial
point for $\boldsymbol{r}$ in the algorithm, which makes it easier
to find a good solution for the non-convex estimation problem \cite{Huangzhe_JRC2}.

Based on the dynamic position grid $\boldsymbol{r}$, we define a
sparse basis $\mathbf{A}\left(\boldsymbol{r}\right)\in\mathbb{C}^{M\times Q}$
as
\begin{equation}
\mathbf{A}\left(\boldsymbol{r}\right)\triangleq\left[\boldsymbol{a}\left(\theta_{1},\phi_{1}\right),\ldots,\boldsymbol{a}\left(\theta_{Q},\phi_{Q}\right)\right]\in\mathbb{C}^{M\times Q}.
\end{equation}
Then, the sparse representation of the NLoS channel vector on the
$n\textrm{-th}$ subcarrier is given by
\begin{equation}
\boldsymbol{h}_{k,n}^{\textrm{NL}}=\mathbf{A}\left(\boldsymbol{r}\right)\mathbf{D}_{k,n}\left(\boldsymbol{r}\right)\boldsymbol{\alpha}_{k},\label{eq:LDhNL}
\end{equation}
where $\mathbf{D}_{k,n}\left(\boldsymbol{r}\right)$ is a diagonal
matrix with the $q\textrm{-th}$ diagonal elements being $e^{-j2\pi nf_{0}\left(\tau_{o,k}+\tau\left(\boldsymbol{r}_{q},\boldsymbol{p}_{u,k}\right)\right)}$,
$\boldsymbol{\alpha}_{k}\in\mathbb{C}^{Q\times1}$ is called the location-domain
NLoS sparse channel vector. $\boldsymbol{\alpha}_{k}$ only has a
few non-zero elements corresponding to the positions of scatterers
in $\mathcal{L}_{k}$. Specifically, the $q\textrm{-th}$ element
of $\boldsymbol{\alpha}_{k}$, denoted by $\alpha_{k,q}$, represents
the complex channel gain of the channel path with the corresponding
scatterer lying in the position $\boldsymbol{r}_{q}$.

\subsection{Three-layer BGG Prior Model}

In practice, the MU channels exhibit certain joint sparsity as explained
in Section \ref{sec:Uplink-System-Model}, i.e., the sets $\mathcal{L}_{k}$'s
have intersections. In this subsection, we shall introduce a three-layer
BGG prior model to capture the joint sparsity of the location-domain
MU channels. Specifically, let $\boldsymbol{\rho}_{k}\triangleq\left[\rho_{k,1},\ldots,\rho_{k,Q}\right]^{T}$
represent the precision vector of $\boldsymbol{\alpha}_{k}$, where
$1/\rho_{k,q}$ is the variance of $\alpha_{k,q}$. Let $\boldsymbol{s}_{k}\triangleq\left[s_{k,1},\ldots,s_{k,Q}\right]^{T}\in\left\{ 0,1\right\} ^{Q}$
represent the support vector of $\boldsymbol{\alpha}_{k}$. If there
is a scatterer of user $k$ around the $q\textrm{-th}$ position grid
$\boldsymbol{r}_{q}$, we have $s_{k,q}=1$ and $\alpha_{k,q}$ is
non-zero. Otherwise, we have $s_{k,q}=0$ and $\alpha_{k,q}=0$. Then,
we introduce a joint support vector $\overline{\boldsymbol{s}}\triangleq\left[\overline{s}_{1},\ldots,\overline{s}_{Q}\right]^{T}\in\left\{ 0,1\right\} ^{Q}$
to represent the union of the positions of all scatterers in $\mathcal{L}$,
i.e., 
\[
\overline{s}_{q}=s_{1,q}\lor s_{2,q}\lor\cdots\lor s_{K,q},
\]
where $\lor$ is the binary logical OR operator. In the proposed three-layer
model, the joint distribution of $\left\{ \boldsymbol{\alpha}_{k},\boldsymbol{\rho}_{k},\boldsymbol{s}_{k},\overline{\boldsymbol{s}}\mid\forall k\right\} $
is represented as
\begin{align*}
 & p\left(\left\{ \boldsymbol{\alpha}_{k},\boldsymbol{\rho}_{k},\boldsymbol{s}_{k},\overline{\boldsymbol{s}}\right\} \right)\\
= & \underbrace{p\left(\overline{\boldsymbol{s}}\right)\prod_{k=1}^{K}p\left(\boldsymbol{s}_{k}\mid\overline{\boldsymbol{s}}\right)}_{\textrm{Support}}\underbrace{p\left(\boldsymbol{\rho}_{k}\mid\boldsymbol{s}_{k}\right)}_{\textrm{Precision}}\underbrace{p\left(\boldsymbol{\alpha}_{k}\mid\boldsymbol{\rho}_{k}\right)}_{\textrm{Sparse\ signal}}.
\end{align*}
A similar three-layer model has been considered in \cite{LiuAn_directloc_vehicles,LiuAn_CE_Turbo_VBI}
and is shown to be more flexible to capture the structured sparsity
of realistic channels.

Conditioned on $\boldsymbol{\rho}_{k}$, the sparse channel $\boldsymbol{\alpha}_{k}$
is assumed to have independent complex Gaussian distributions with
zero mean and variance $1/\boldsymbol{\rho}_{k}$, i.e.,
\begin{equation}
\begin{aligned}p\left(\boldsymbol{\alpha}_{k}\mid\boldsymbol{\rho}_{k}\right) & =\prod_{q}\mathcal{CN}\left(\alpha_{k,q};0,\frac{1}{\rho_{k,q}}\right).\end{aligned}
\end{equation}
The conditional distribution $p\left(\boldsymbol{\rho}_{k}\mid\boldsymbol{s}_{k}\right)$
is given by
\begin{equation}
\begin{aligned}p\left(\boldsymbol{\rho}_{k}\mid\boldsymbol{s}_{k}\right)= & \prod_{q}\left(\delta\left(s_{k,q}-1\right)\Gamma\left(\rho_{k,q};a,b\right)\right.\\
+ & \left.\delta\left(s_{k,q}\right)\Gamma\left(\rho_{k,q};\overline{a},\overline{b}\right)\right),
\end{aligned}
\end{equation}
where $\delta\left(\cdot\right)$ is the Dirac Delta function, $\Gamma\left(\rho;a,b\right)$
is a Gamma hyper-prior with shape parameter $a$ and rate parameter
$b$. When $s_{k,q}=1$, $\alpha_{k,q}$ is a non-zero element and
the corresponding variance $1/\rho_{k,q}$ is $\mathcal{O}\left(1\right)$.
In this case, $a$ and $b$ should be chosen to satisfy $\tfrac{a}{b}=\mathbb{E}\left(\rho_{k,q}\right)=\mathcal{O}\left(1\right)$.
When $s_{k,q}=0$, $\alpha_{k,q}$ is a zero element and the corresponding
variance $1/\rho_{k,q}$ is close to zero. In this case, $\overline{a}$
and $\overline{b}$ should be chosen to satisfy $\tfrac{\overline{a}}{\overline{b}}=\mathbb{E}\left(\rho_{k,q}\right)\gg1$.
A typical value is $a=1$, $b=1$, $\overline{a}=1$, and $\overline{b}=10^{-5}$
\cite{LiuAn_CE_Turbo_VBI}.

The conditional probability for the support vector is given by $p\left(\boldsymbol{s}_{k}\mid\overline{\boldsymbol{s}}\right)=\prod_{q}p\left(s_{k,q}\mid\overline{s}_{q}\right)$
with
\begin{align}
p\left(s_{k,q}\mid\overline{s}_{q}\right) & =\delta\left(\overline{s}_{q}\right)\delta\left(s_{k,q}\right)+\delta\left(\overline{s}_{q}-1\right)\\
\times & \left(\delta\left(s_{k,q}-1\right)\lambda{}_{k,q}+\delta\left(s_{k,q}\right)\left(1-\lambda{}_{k,q}\right)\right),\nonumber 
\end{align}
where $\lambda{}_{k,q}$ represents the probability of $s_{k,q}=1$
conditioned on $\overline{s}_{q}=1$, depicting the degree of overlap
between scatterers of individual users and global scatterers. The
joint support vector $\overline{\boldsymbol{s}}$ is modeled as an
independent prior: $p\left(\overline{\boldsymbol{s}}\right)=\prod_{q}p\left(\overline{s}_{q}\right)$
with $p\left(\overline{s}_{q}=1\right)=\lambda_{q}$.

To automatically learn the noise precision, we assume a gamma distribution
with shape parameter $c$ and rate parameter $d$ as the prior for
$\gamma$, i.e., $p\left(\gamma\right)=\Gamma\left(\gamma;c,d\right).$
Let $\rho_{k,0}$ and $s_{k,0}$ denote the precision and support
of the LoS channel gain $\alpha_{k,0}$, respectively. Then, the sparse
prior model for the LoS channel $p\left(\alpha_{k,0},\rho_{k,0},s_{k,0}\right)$
is similar except that there is no joint support, and the probability
of $s_{k,0}=1$ is $\lambda_{k,0}$.

\subsection{Bilinear Sparse Bayesian Inference with Uncertain Parameters}

For convenience, define pilot matrix $\mathbf{U}_{t_{p}}=\left[\mathbf{U}_{t_{p},1},...,\mathbf{U}_{t_{p},K}\right]$
and data matrix $\mathbf{X}_{t_{d}}=\left[\mathbf{X}_{t_{d},1},...,\mathbf{X}_{t_{d},K}\right]$
with
\begin{align*}
\mathbf{U}_{t_{p},k} & =\text{BlockDiag}\left(u_{t_{p},k,1}\mathbf{I}_{M},...,u_{t_{p},k,N}\mathbf{I}_{M}\right),\\
\mathbf{X}_{t_{d},k} & =\text{BlockDiag}\left(x_{t_{d},k,1}\mathbf{I}_{M},...,x_{t_{d},k,N}\mathbf{I}_{M}\right),
\end{align*}
where $u_{t_{p},k,n}=0,\forall n\notin\mathcal{N}_{t_{p},k}^{p}$
and $x_{t_{d},k,n}=0,\forall n\notin\mathcal{N}_{t_{d},k}^{d}$. Let
$\boldsymbol{\xi}\triangleq\left\{ \boldsymbol{r},\boldsymbol{p}_{u},\tau_{o}\right\} $
with $\boldsymbol{p}_{u}=\left\{ \boldsymbol{p}_{u,1},...,\boldsymbol{p}_{u,k}\right\} $
denote the collection of sensing parameters. Moreover, define sensing
matrix $\boldsymbol{\Phi}\left(\boldsymbol{\xi}\right)\triangleq\mathrm{\textrm{BlockDiag}}\left(\boldsymbol{\Phi}_{1}\left(\boldsymbol{\xi}\right),...,\boldsymbol{\Phi}_{K}\left(\boldsymbol{\xi}\right)\right)$,
where
\begin{equation}
\begin{aligned}\boldsymbol{\Phi}_{k}\left(\boldsymbol{\xi}\right) & \triangleq\left[\boldsymbol{\Phi}_{k,n}^{0},\boldsymbol{\Phi}_{k,n}^{1}\right]_{n\in\mathcal{N}}\in\mathbb{C}^{MN\times\left(Q+1\right)},\end{aligned}
\end{equation}
with
\begin{align*}
\boldsymbol{\Phi}_{k,n}^{0} & =\left[e^{-j2\pi nf_{0}\left(\tau_{o,k}+\tau_{0}\left(\boldsymbol{p}_{u,k}\right)\right)}\boldsymbol{a}\left(\theta_{u,k},\phi_{u,k}\right)\right],\\
\boldsymbol{\Phi}_{k,n}^{1} & =\left[\mathbf{A}\left(\boldsymbol{r}\right)\mathbf{D}_{k,n}\left(\boldsymbol{r}\right)\right].
\end{align*}
Then, using the location-domain sparse representation in (\ref{eq:LDhNL}),
the received pilot and data signals on all available subcarriers can
be expressed as
\begin{subequations}
\begin{align}
\boldsymbol{y}_{t_{p}}^{p} & =\mathbf{U}_{t_{p}}\boldsymbol{\Phi}\left(\boldsymbol{\xi}\right)\boldsymbol{\alpha}+\boldsymbol{z}^{p},\forall t_{p},\label{eq:yr}\\
\boldsymbol{y}_{t_{d}}^{d} & =\mathbf{X}_{t_{d}}\boldsymbol{\Phi}\left(\boldsymbol{\xi}\right)\boldsymbol{\alpha}+\boldsymbol{z}^{d},\forall t_{d},\label{eq:yc}
\end{align}
\end{subequations}
respectively, where $\boldsymbol{y}_{t_{p}}^{p}\triangleq\left[\boldsymbol{y}_{t_{p},n}^{p}\right]_{n\in\mathcal{N}}\in\mathbb{C}^{MN\times1}$,
$\boldsymbol{y}_{t_{d}}^{d}\triangleq\left[\boldsymbol{y}_{t_{d},n}^{d}\right]_{n\in\mathcal{N}}\in\mathbb{C}^{MN\times1}$,
$\boldsymbol{\alpha}=\left[\overline{\boldsymbol{\alpha}}_{1}^{T},...,\overline{\boldsymbol{\alpha}}_{K}^{T}\right]^{T}$
with $\overline{\boldsymbol{\alpha}}_{k}=\left[\alpha_{k,0},\boldsymbol{\alpha}_{k}^{T}\right]^{T}$,
$\boldsymbol{z}_{t_{p}}^{p}\triangleq\left[\boldsymbol{z}_{t_{p},n}^{p}\right]_{n\in\mathcal{N}}\in\mathbb{C}^{MN\times1}$,
and $\boldsymbol{z}_{t_{d}}^{d}\triangleq\left[\boldsymbol{z}_{t_{d},n}^{d}\right]_{n\in\mathcal{N}}\in\mathbb{C}^{MN\times1}$.

For given sensing parameters $\boldsymbol{\xi}$ and data $\boldsymbol{x}_{t_{d}}=\left[\boldsymbol{x}_{t_{d},1}^{T},...,\boldsymbol{x}_{t_{d},N}^{T}\right]^{T}\in\mathbb{C}^{\sum_{n=1}^{N}\left|\mathcal{K}_{t_{d},n}^{d}\right|\times1}$
with $\boldsymbol{x}_{t_{d},n}=\left[x_{t_{d},k,n}\right]_{k\in\mathcal{K}_{t_{d},n}^{d}}\in\mathbb{C}^{\left|\mathcal{K}_{t_{d},n}^{d}\right|\times1},\forall t_{d},n$,
(\ref{eq:yr}) and (\ref{eq:yc}) form a linear observation model
for the estimation of channel $\boldsymbol{\alpha}$. For given channel
$\boldsymbol{\alpha}$ and sensing parameters $\boldsymbol{\xi}$,
(\ref{eq:yc}) also forms a linear observation model for data recovery
as
\begin{align*}
\boldsymbol{y}_{t_{d},n}^{d} & =\mathbf{\Psi}_{n}\left(\boldsymbol{\xi}\right)\boldsymbol{x}_{t_{d},n}+\boldsymbol{z}_{t_{d},n}^{d},\forall t_{d},n.
\end{align*}
where 
\begin{equation}
\mathbf{\Psi}_{n}\left(\boldsymbol{\xi}\right)\triangleq\left[\left[\left(\boldsymbol{\Phi}_{k,n}\left(\boldsymbol{\xi}\right)\overline{\boldsymbol{\alpha}}_{k}\right)^{T}\right]_{k\in\mathcal{K}_{t_{d},n}^{d}}\right]^{T}\in\mathbb{C}^{M\times\left|\mathcal{K}_{t_{d},n}^{d}\right|}
\end{equation}
 with $\begin{aligned}\boldsymbol{\Phi}_{k,n}\left(\boldsymbol{\xi}\right) & \triangleq\left[\boldsymbol{\Phi}_{k,n}^{0},\boldsymbol{\Phi}_{k,n}^{1}\right]\in\mathbb{C}^{M\times\left(Q+1\right)}.\end{aligned}
$

Therefore, (\ref{eq:yr}) and (\ref{eq:yc}) is a bilinear observation
model w.r.t. $\boldsymbol{\alpha}$ and $\boldsymbol{x}_{t_{d}}$'s
with uncertain sensing parameters $\boldsymbol{\xi}$.

To simplify the notation, the precision vector and the support vector
of $\boldsymbol{\alpha}$ are respectively defined as
\begin{align*}
\boldsymbol{\rho} & \triangleq\left[\rho_{1,0},\boldsymbol{\rho}_{1}^{T},...,\rho_{K,0},\boldsymbol{\rho}_{K}^{T}\right]^{T},\\
\boldsymbol{s} & \triangleq\left[s_{1,0},\boldsymbol{s}_{1}^{T},...,s_{K,0},\boldsymbol{s}_{K}^{T}\right]^{T}.
\end{align*}
Our primary goal is to estimate the channel vector $\boldsymbol{\alpha}$,
the data vector $\boldsymbol{x}\triangleq\left[\left(\boldsymbol{x}_{1}\right)^{T},...,\left(\boldsymbol{x}_{T_{d}}\right)^{T}\right]^{T}$,
the support vector $\boldsymbol{s}$, and the uncertain parameters
$\boldsymbol{\xi}$ given observation $\boldsymbol{y}\triangleq\left[\left(\boldsymbol{y}_{1}^{p}\right)^{T},...,\left(\boldsymbol{y}_{T_{p}}^{p}\right)^{T},\left(\boldsymbol{y}_{1}^{d}\right)^{T},...,\left(\boldsymbol{y}_{T_{d}}^{d}\right)^{T}\right]^{T}$.
To be specific, for given $\boldsymbol{\xi}$, we aim at computing
the conditional marginal posteriors, i.e., $p\left(\boldsymbol{\alpha}\mid\boldsymbol{y};\boldsymbol{\xi}\right)$,
$p\left(\boldsymbol{x}\mid\boldsymbol{y};\boldsymbol{\xi}\right)$,
and $p\left(s_{i}\mid\boldsymbol{y};\boldsymbol{\xi}\right),\forall i$.
On the other hand, the uncertain parameters $\boldsymbol{\xi}$ are
obtained by the MAP estimator as follows:
\begin{equation}
\begin{aligned}\boldsymbol{\xi}^{\ast} & =\underset{\boldsymbol{\xi}}{\arg\max}\ln p\left(\boldsymbol{y},\boldsymbol{\xi}\right)\\
 & =\underset{\boldsymbol{\xi}}{\arg\max}\ln\int_{\overline{\boldsymbol{v}}}p\left(\boldsymbol{y},\overline{\boldsymbol{v}};\boldsymbol{\xi}\right)p\left(\boldsymbol{\xi}\right),
\end{aligned}
\label{eq:M-step}
\end{equation}
where $\overline{\boldsymbol{v}}\triangleq\left\{ \boldsymbol{\alpha},\boldsymbol{x},\boldsymbol{\rho},\boldsymbol{s},\overline{\boldsymbol{s}},\gamma\right\} $
denotes the collection of all variables, and $p\left(\boldsymbol{\xi}\right)$
denotes the known prior distribution of $\boldsymbol{\xi}$. Once
we obtain the MAP estimate of $\boldsymbol{\xi}^{\ast}$, we can obtain
the minimum mean square error (MMSE) estimate of $\boldsymbol{\alpha}$
as $\boldsymbol{\alpha}^{*}=\int_{\boldsymbol{\alpha}}\boldsymbol{\alpha}p\left(\boldsymbol{\alpha}\mid\boldsymbol{y};\boldsymbol{\xi}^{\ast}\right)$,
and the MAP estimate of $\boldsymbol{s}$ as $s_{k,q}^{\ast}=\arg\max_{s_{k,q}}p\left(s_{k,q}\mid\boldsymbol{y};\boldsymbol{\xi}^{\ast}\right),\forall k,q$.
Moreover, the marginal posteriors $p\left(\boldsymbol{x}\mid\boldsymbol{y};\boldsymbol{\xi}\right)$
for the data can be used as soft information for digital demodulation
and channel decoding.

However, it is exceedingly challenging to calculate the above conditional
marginal posteriors precisely due to the looped factor graph. In the
following section, we present the EM-Turbo-BiSVBI algorithm to solve
this problem efficiently.

\section{EM-Turbo-BiSVBI Algorithm\label{sec:SEA-Turbo-SBI-Algorithm}}

\subsection{Outline of the EM-Turbo-BiSVBI Algorithm}

\begin{figure}[t]
\begin{centering}
\textsf{\includegraphics[width=80mm]{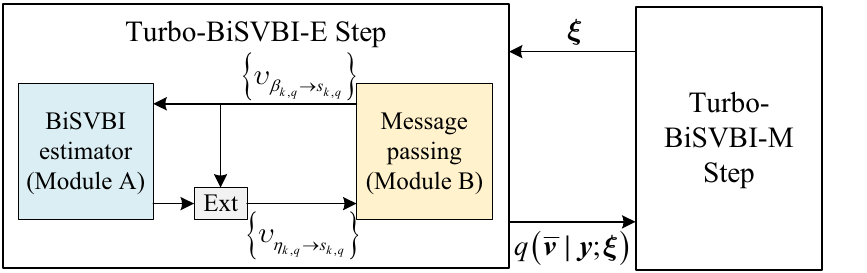}}
\par\end{centering}
\caption{\label{fig:Turbo-BISVBI}Framework of the EM-Turbo-BiSVBI algorithm.}
\end{figure}

As illustrated in Fig. \ref{fig:Turbo-BISVBI}, the EM-Turbo-BiSVBI
algorithm iterates between the following two major steps until convergence.

\begin{itemize}
\item \textbf{Turbo-BiSVBI-E Step:} For given $\boldsymbol{\xi}^{\left(t\right)}$
in the $t\textrm{-th}$ iteration, apply the turbo approach to combine
the BiSVBI method and the message passing method to calculate approximate
marginal posteriors, denoted by $q\left(\overline{\boldsymbol{v}}\mid\boldsymbol{y};\boldsymbol{\xi}^{\left(t\right)}\right)$.
\item \textbf{Turbo-BiSVBI-M Step:} Apply the EM method to construct a surrogate
function for the MAP objective $\ln p\left(\boldsymbol{y},\boldsymbol{\xi}\right)$
based on $q\left(\overline{\boldsymbol{v}}\mid\boldsymbol{y};\boldsymbol{\xi}^{\left(t\right)}\right)$,
then maximize the surrogate function with respect to $\boldsymbol{\xi}$.
\end{itemize}
In the following, we will elaborate the Turbo-BiSVBI-E and M Steps.
\begin{table}[t]
\caption{\label{tab:Factors_Table}Factors, Distributions, and Functional forms
in Fig. \ref{fig:two_subgraph}.}

\centering{}%
\begin{tabular}{|c|c|c|}
\hline 
{\small Factor} & {\small Distribution} & {\small Functional form}\tabularnewline
\hline 
\hline 
{\small$\begin{array}{c}
f_{t_{p}}^{p}\\
f_{t_{d}}^{d}
\end{array}$} & {\small$\begin{array}{c}
p\left(\boldsymbol{y}_{t_{p}}^{p}\mid\boldsymbol{\alpha}\right)\\
p\left(\boldsymbol{y}_{t_{d}}^{d}\mid\boldsymbol{x}_{t_{d}},\boldsymbol{\alpha}\right)
\end{array}$} & {\small$\begin{array}{c}
\mathcal{CN}\left(\boldsymbol{y}_{t_{p}}^{p};\mathbf{U}_{t_{p}}\boldsymbol{\Phi}\left(\boldsymbol{\xi}\right)\boldsymbol{\alpha},1/\gamma\right)\\
\mathcal{CN}\left(\boldsymbol{y}_{t_{d}}^{d};\mathbf{X}_{t_{d}}\boldsymbol{\Phi}\left(\boldsymbol{\xi}\right)\boldsymbol{\alpha},1/\gamma\right)
\end{array}$}\tabularnewline
\hline 
{\small$g_{t_{d}}^{d}$} & {\small$\prod_{k,n}p\left(x_{t_{d},k,n}\right)$} & {\small$\prod_{k,n}\mathcal{CN}\left(x_{t_{d},k,n};0,\sigma_{t_{d},k,n}^{2}\right)$}\tabularnewline
\hline 
{\small$g_{k,q}$} & {\small$p\left(\alpha_{k,q}\mid\rho_{k,q}\right)$} & {\small$\mathcal{CN}\left(\alpha_{k,q};0,1/\rho_{k,q}\right)$}\tabularnewline
\hline 
{\small$\eta_{k,q}$} & {\small$p\left(\rho_{k,q}\mid s_{k,q}\right)$} & {\small$\begin{cases}
\Gamma\left(\rho_{k,q};a,b\right), & s_{k,q}=1\\
\Gamma\left(\rho_{k,q};\overline{a},\overline{b}\right), & s_{k,q}=0
\end{cases}$}\tabularnewline
\hline 
{\small$\beta_{k,q}$} & {\small$p\left(s_{k,q}\mid\overline{s}_{q}\right)$} & {\small$\begin{cases}
p\left(s_{k,q}=1\right)=\lambda{}_{k,q}, & \overline{s}_{q}=1\\
s_{k,q}=0, & \overline{s}_{q}=0
\end{cases}$}\tabularnewline
\hline 
{\small$\kappa_{q}$} & {\small$p\left(\overline{s}_{q}\right)$} & {\small$p\left(\overline{s}_{q}=1\right)=\lambda{}_{q}$}\tabularnewline
\hline 
\end{tabular}
\end{table}

\subsection{Turbo-BiSVBI-E Step}

\begin{figure}[t]
\centering{}\includegraphics[width=90mm]{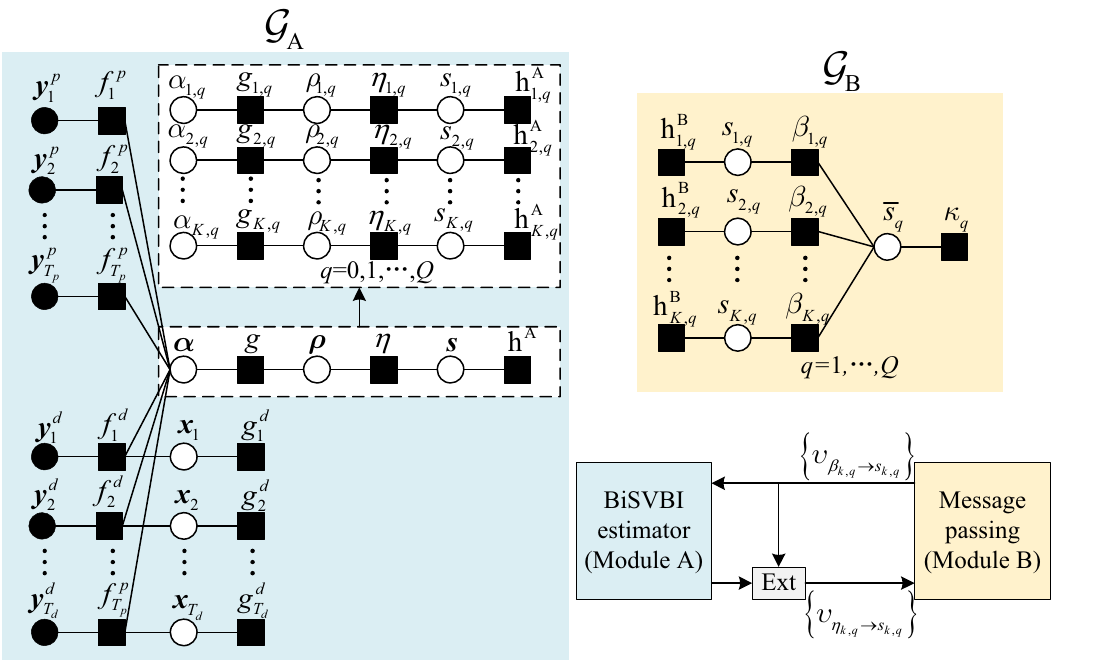}\caption{\label{fig:two_subgraph}Module A and Module B of the Turbo-BiSVBI-E
Step and messages flow between two modules.}
\end{figure}

In the E step, the factor graph of the joint distribution $p\left(\boldsymbol{y},\overline{\boldsymbol{v}};\boldsymbol{\xi}\right)$
is partitioned into two subgraphs, denoted by $\mathcal{G}_{\mathrm{A}}$
and $\mathcal{G}_{\mathrm{B}}$, as shown in Fig. \ref{fig:two_subgraph},
where the expressions of each factor node are listed in Table \ref{tab:Factors_Table}.
To be more specific, $\mathcal{G}_{\mathrm{A}}$ describes the structure
of the bilinear observation model with an independent BGG prior, while
$\mathcal{G}_{\mathrm{B}}$ describes the more complicated sparse
structure of the support vector $p\left(\boldsymbol{s},\bar{\boldsymbol{s}}\right)$
with effective binary observations. Correspondingly, two modules called
Module A and B are designed to perform Bayesian inference over the
two subgraphs.

Specifically, the extrinsic messages from Module A to B are $\left\{ \upsilon_{\eta_{k,q}\rightarrow s_{k,q}}\right\} $,
while the extrinsic messages from Module B to A are $\left\{ \upsilon_{\beta_{k,q}\rightarrow s_{k,q}}\right\} $,
as shown in Fig. \ref{fig:two_subgraph}. For convenience, we define
two turbo-iteration factor nodes to represent the extrinsic messages
in $\mathcal{G}_{\mathrm{A}}$ and $\mathcal{G}_{\mathrm{B}}$:
\begin{equation}
\begin{aligned}\textrm{h}_{k,q}^{\text{A}} & \triangleq\upsilon_{\beta_{k,q}\rightarrow s_{k,q}}\left(s_{k,q}\right) & \textrm{h}_{k,q}^{\text{B}} & \triangleq\upsilon_{\eta_{k,q}\rightarrow s_{k,q}}\left(s_{k,q}\right)\end{aligned}
,\forall k,q.\label{eq:turbo-iteration-factor}
\end{equation}
For each turbo iteration, Module A treats $\textrm{h}_{k,q}^{\text{A}}$
as the prior and performs the BiSVBI to calculate the approximate
conditional posteriors $q\left(s_{k,q}\right)$'s. Then, Module A
passes extrinsic messages to Module B by subtracting the prior information
from posterior information in the log-probability domain, i.e.,
\begin{equation}
\textrm{h}_{k,q}^{\text{B}}\triangleq\upsilon_{\eta_{k,q}\rightarrow s_{k,q}}\left(s_{k,q}\right)\propto q\left(s_{k,q}\right)/\textrm{h}_{k,q}^{\text{A}}.\label{eq:extricsic message}
\end{equation}
Similarly, Module B treats $\textrm{h}_{k,q}^{\text{B}}$ as the binary
observation information and performs message passing over $\mathcal{G_{B}}$
and passes the extrinsic messages $\textrm{h}_{k,q}^{\text{A}}\triangleq\upsilon_{\beta_{k,q}\rightarrow s_{k,q}}\left(s_{k,q}\right)$
to Module A. The two modules iterate until they reach a point of convergence. 

Since Module B is a simply tree graph and the associated marginal
posteriors can be easily calculated via sum-product message passing
\cite{Ksch_sum-product}, we will only elaborate on Module A (BiSVBI)
in the next subsection. 

\subsection{BiSVBI Module (Module A)}

\subsubsection{Formulation of BiSVBI}

The BiSVBI module generalizes the SC-VBI in \cite{Liu_SCVBI} from
a linear observation model to a bilinear model to calculate the approximate
marginal posterior $q\left(\boldsymbol{v}\right)$ for fixed sensing
parameters $\boldsymbol{\xi}$, where $\boldsymbol{v}\triangleq\left\{ \boldsymbol{\alpha},\boldsymbol{x},\boldsymbol{\rho},\boldsymbol{s},\gamma\right\} $.
To be more specific, the turbo-iteration factor nodes $\textrm{h}_{k,q}^{\text{A}}\left(s_{k,q}\right),\forall k,q$
incorporate the prior information from Module B. Therefore, the following
prior distribution is assumed in the BiSVBI module:
\begin{align}
\hat{p}\left(\boldsymbol{v}\right) & =\hat{p}\left(\boldsymbol{s}\right)p\left(\boldsymbol{\rho}\mid\boldsymbol{s}\right)p\left(\boldsymbol{\alpha}\mid\boldsymbol{\rho}\right)p\left(\boldsymbol{x}\right)p\left(\gamma\right),\nonumber \\
\hat{p}\left(\boldsymbol{s}\right) & =\prod_{k=1}^{K}\prod_{q=1}^{Q}\textrm{h}_{k,q}^{\text{A}}\left(s_{k,q}\right).\label{eq:VBIpri}
\end{align}
Since the sensing parameter $\boldsymbol{\xi}$ is fixed in BiSVBI,
we omit $\boldsymbol{\xi}$ in all notations in this subsection. 

For convenience, we use $\boldsymbol{v}^{k},k\in\mathcal{H}\triangleq\left\{ 1,2,...,5\right\} $
to denote an individual variable in $\boldsymbol{v}$. Denote $\hat{p}\left(\boldsymbol{v}\mid\boldsymbol{y}\right)$
as the posterior distribution of $\boldsymbol{v}$ with the prior
$\hat{p}\left(\boldsymbol{v}\right)$ in (\ref{eq:VBIpri}). Based
on the VBI method, the approximate marginal posterior could be calculated
by minimizing the KLD between $\hat{p}\left(\boldsymbol{v}\mid\boldsymbol{y}\right)$
and $q\left(\boldsymbol{v}\right)$, subject to a factorized form
constraint as \cite{Tzikas_VBI}
\begin{eqnarray}
\mathcal{\mathscr{A}}_{\mathrm{VBI}}: & q^{*}\left(\boldsymbol{v}\right) & =\arg\min_{q\left(\boldsymbol{v}\right)}\int q\left(\boldsymbol{v}\right)\ln\frac{q\left(\boldsymbol{v}\right)}{\hat{p}\left(\boldsymbol{v}\mid\boldsymbol{y}\right)}\textrm{d}\boldsymbol{v},\label{eq:KLDmin}\\
\mathrm{s.t.} & q\left(\boldsymbol{v}\right) & =\prod_{k\in\mathcal{H}}q\left(\boldsymbol{v}^{k}\right),\int q\left(\boldsymbol{v}^{k}\right)d\boldsymbol{v}^{k}=1,\label{eq:factorconstrain}\\
 & q\left(\boldsymbol{\alpha}\right) & =\mathcal{CN}\left(\boldsymbol{\alpha};\boldsymbol{\mu}_{\alpha},\textrm{Diag}\left(\boldsymbol{\sigma}_{\alpha}^{2}\right)\right),\label{eq:GScon}\\
 & q\left(\boldsymbol{x}\right) & =\mathcal{CN}\left(\boldsymbol{x};\boldsymbol{\mu}_{x},\textrm{Diag}\left(\boldsymbol{\sigma}_{x}^{2}\right)\right).\label{eq:GScon_x}
\end{eqnarray}
Note that we add an additional constraint that $q\left(\boldsymbol{\alpha}\right)$/$q\left(\boldsymbol{x}\right)$
is a Gaussian distribution with mean $\boldsymbol{\mu}_{\alpha}$/$\boldsymbol{\mu}_{x}$
and diagonal covariance matrix $\textrm{Diag}\left(\boldsymbol{\sigma}_{\alpha}^{2}\right)$/$\textrm{Diag}\left(\boldsymbol{\sigma}_{x}^{2}\right)$,
where both the mean $\boldsymbol{\mu}_{\alpha}$/$\boldsymbol{\mu}_{x}$
and variance vector $\boldsymbol{\sigma}_{\alpha}^{2}$/$\boldsymbol{\sigma}_{x}^{2}$
are optimization variables. This additional constraint is imposed
to avoid the high-dimensional matrix inverse for calculating the posterior
covariance matrix. And the high-dimensional matrix inverse for calculating
the posterior mean will be avoided using the subspace constrained
method. Since Problem $\mathcal{\mathscr{A}}_{\mathrm{VBI}}$ is non-convex,
we aim at finding a stationary solution of $\mathcal{\mathscr{A}}_{\mathrm{VBI}}$,
denoted by $q^{*}\left(\boldsymbol{v}\right)$, as defined below.
\begin{defn}
[Stationary Solution]\label{lem:optimality-conditon-1}$q^{*}\left(\boldsymbol{v}\right)=\prod_{k\in\mathcal{H}}q^{*}\left(\boldsymbol{v}^{k}\right)$
is called a stationary solution of Problem $\mathcal{\mathscr{A}}_{\mathrm{VBI}}$
if it satisfies all the constraints in $\mathcal{\mathscr{A}}_{\mathrm{VBI}}$
and $\forall k\in\mathcal{H}$,
\begin{multline*}
q^{*}\left(\boldsymbol{v}^{k}\right)=\\
\arg\min_{q\left(\boldsymbol{v}^{k}\right)}\int\prod_{l\neq k}q^{*}\left(\boldsymbol{v}^{l}\right)q\left(\boldsymbol{v}^{k}\right)\ln\frac{\prod_{l\neq k}q^{*}\left(\boldsymbol{v}^{l}\right)q\left(\boldsymbol{v}^{k}\right)}{\hat{p}\left(\boldsymbol{v}\mid\boldsymbol{y}\right)}.
\end{multline*}
In particular, for $q\left(\boldsymbol{\alpha}\right)$ and $q\left(\boldsymbol{x}\right)$,
the above minimization is over the optimization variables $\boldsymbol{\mu}_{\alpha},\boldsymbol{\sigma}_{\alpha}^{2}$
and $\boldsymbol{\mu}_{x},\boldsymbol{\sigma}_{x}^{2}$.

\end{defn}
In the following, we shall derive the BiSVBI algorithm to find a stationary
solution for $\mathcal{\mathscr{A}}_{\mathrm{VBI}}$.

\subsubsection{Derivation of the BiSVBI Algorithm for $\mathcal{\mathscr{A}}_{\mathrm{VBI}}$}

The BiSVBI algorithm finds a stationary solution of $\mathcal{\mathscr{A}}_{\mathrm{VBI}}$
via alternately optimizing each individual density $q\left(\boldsymbol{v}^{k}\right),k\in\mathcal{H}$.
Specifically, for given $q\left(\boldsymbol{v}^{l}\right),\forall l\neq k$
and $\boldsymbol{v}^{k}\neq\boldsymbol{\alpha}$ or $\boldsymbol{x}$,
the optimal $q\left(\boldsymbol{v}^{k}\right)$ that minimizes the
KLD in $\mathcal{\mathscr{A}}_{\mathrm{VBI}}$ is given by \cite{Tzikas_VBI}
\begin{equation}
q\left(\boldsymbol{v}^{k}\right)\propto\exp\left(\left\langle \ln\hat{p}\left(\boldsymbol{v},\boldsymbol{y}\right)\right\rangle _{\prod_{l\neq k}q\left(\boldsymbol{v}^{l}\right)}\right),\label{eq:optimal_q}
\end{equation}
where $\left\langle f\left(x\right)\right\rangle _{q\left(x\right)}=\int f\left(x\right)q\left(x\right)dx$
and $\hat{p}\left(\boldsymbol{v},\boldsymbol{y}\right)$ is the joint
distribution with prior $\hat{p}\left(\boldsymbol{v}\right)$. However,
when $\boldsymbol{v}^{k}=\boldsymbol{\alpha}$ or $\boldsymbol{x}$,
the optimal $q\left(\boldsymbol{\alpha}\right)$ or $q\left(\boldsymbol{x}\right)$
is no longer given by (\ref{eq:optimal_q}) due to the independence
constraint, and it is characterized by the following lemma. Note that
the expectation $\left\langle f\left(\boldsymbol{v}^{k}\right)\right\rangle _{q\left(\boldsymbol{v}^{k}\right)}$
w.r.t. its own approximate posterior will be simplified as $\left\langle f\left(\boldsymbol{v}^{k}\right)\right\rangle $.
\begin{lem}
\label{lem:ICVBI}For given $q\left(\boldsymbol{v}^{k}\right),\boldsymbol{v}^{k}\neq\boldsymbol{\alpha}$,
the optimal posterior mean $\boldsymbol{\mu}_{\alpha}$ and variance
vector $\boldsymbol{\sigma}_{\alpha}^{2}$ that minimizes the KLD
subject to $q\left(\boldsymbol{\alpha}\right)=\mathcal{CN}\left(\boldsymbol{\alpha};\boldsymbol{\mu}_{\alpha},\textrm{Diag}\left(\boldsymbol{\sigma}_{\alpha}^{2}\right)\right)$
are given by
\begin{align}
\boldsymbol{\mu}_{\alpha} & =\underset{\boldsymbol{u}_{\alpha}}{\arg\min}\ \varphi_{\alpha}\left(\boldsymbol{u}_{\alpha}\right)\triangleq\boldsymbol{u}_{\alpha}^{H}\mathbf{W}_{\alpha}\boldsymbol{u}_{\alpha}-2\mathfrak{Re}\left\{ \boldsymbol{u}_{\alpha}^{H}\boldsymbol{b}_{\alpha}\right\} ,\nonumber \\
\boldsymbol{\sigma}_{\alpha}^{2} & =\left[\mathrm{W}_{\alpha,1,0}^{-1},...,\mathrm{W}_{\alpha,1,Q}^{-1},...,\mathrm{W}_{\alpha,K,0}^{-1},...,\mathrm{W}_{\alpha,K,Q}^{-1}\right]^{T}.\label{eq:muasiga}
\end{align}
 where $\mathbf{W}_{\alpha}=\textrm{Diag}\left(\left\langle \boldsymbol{\rho}\right\rangle \right)+\left\langle \gamma\right\rangle \left\langle \mathbf{A}_{\alpha}^{H}\mathbf{A}_{\alpha}\right\rangle $,
$\mathrm{W}_{\alpha,k,q}$ is the $\left[\left(k-1\right)\left(Q+1\right)+q+1\right]$-th
diagonal element of $\mathbf{W}_{\alpha}$, and $\boldsymbol{b}_{\alpha}=\left\langle \gamma\right\rangle \left\langle \mathbf{A}_{\alpha}^{H}\right\rangle \boldsymbol{y}$
with $\mathbf{A}_{\alpha}=\left[\mathbf{A}_{\alpha}^{p};\mathbf{A}_{\alpha}^{d}\right]$,
\begin{align*}
\mathbf{A}_{\alpha}^{p}= & \left[\mathbf{U}_{1}\boldsymbol{\Phi};\ldots;\mathbf{U}_{T_{p}}\boldsymbol{\Phi}\right],\\
\mathbf{A}_{\alpha}^{d}= & \left[\mathbf{X}_{1}\boldsymbol{\Phi};\ldots;\mathbf{X}_{T_{d}}\boldsymbol{\Phi}\right].
\end{align*}
Similarly, for given $q\left(\boldsymbol{v}^{k}\right),\boldsymbol{v}^{k}\neq\boldsymbol{x}$,
the optimal posterior mean $\boldsymbol{\mu}_{x}$ and variance vector
$\boldsymbol{\sigma}_{x}^{2}$ are given by
\begin{align}
\boldsymbol{\mu}_{x,t_{d},n} & =\underset{\boldsymbol{u}_{x,t_{d},n}}{\arg\min}\ \varphi_{x,t_{d},n}\left(\boldsymbol{u}_{x,t_{d},n}\right)\nonumber \\
 & \triangleq\boldsymbol{u}_{x,t_{d},n}^{H}\mathbf{W}_{x,t_{d},n}\boldsymbol{u}_{x,t_{d},n}-2\mathfrak{Re}\left\{ \boldsymbol{u}_{x,t_{d},n}^{H}\boldsymbol{b}_{x,t_{d},n}\right\} ,\nonumber \\
\boldsymbol{\sigma}_{x,t_{d},n}^{2} & =\left[\mathrm{W}_{x,t_{d},k,n}^{-1}\right]_{k\in\mathcal{K}_{t_{d},n}^{d}},\forall t_{d},n.\label{eq:muxsigx}
\end{align}
where $\boldsymbol{\mu}_{x,t_{d},n}$ and $\boldsymbol{\sigma}_{x,t_{d},n}^{2}$
are the posterior mean and variance corresponding to $\boldsymbol{x}_{t_{d},n}$,
$\mathbf{W}_{x,t_{d},n}\triangleq\textrm{Diag}\left(\boldsymbol{\sigma}_{t_{d},n}^{2}\right)+\left\langle \gamma\right\rangle \left\langle \mathbf{\Psi}_{n}^{H}\mathbf{\Psi}_{n}\right\rangle $
with $\boldsymbol{\sigma}_{t_{d},n}^{2}=\left[\sigma_{k,n}^{2}\right]_{k\in\mathcal{K}_{t_{d},n}^{d}}$,
$\boldsymbol{b}_{x,t_{d},n}=\left\langle \gamma\right\rangle \left\langle \mathbf{\Psi}_{n}^{H}\right\rangle \boldsymbol{y}_{t_{d},n}^{d}$,
and $\left[\mathrm{W}_{x,t_{d},k,n}\right]_{k\in\mathcal{K}_{t_{d},n}^{d}}=\textrm{Diag}\left(\mathbf{W}_{x,t_{d},n}\right)$.
\end{lem}
Please refer to Appendix A for the detailed proof. Based on \eqref{eq:optimal_q}
and Lemma \ref{lem:ICVBI}, the update equations of all variables
are given below.

\paragraph{Update Equation for $q\left(\boldsymbol{\alpha}\right)$}

From Lemma \ref{lem:ICVBI}, $q\left(\boldsymbol{\alpha}\right)$
can be derived as
\begin{equation}
q\left(\boldsymbol{\alpha}\right)=\mathcal{CN}\left(\boldsymbol{\mathbf{\alpha}};\boldsymbol{\mu}_{\alpha},\boldsymbol{\sigma}_{\alpha}^{2}\right),\label{eq:poster_x}
\end{equation}
where $\boldsymbol{\mu}_{\alpha}=\mathbf{W}_{\alpha}^{-1}\boldsymbol{b}_{\alpha}$
and $\boldsymbol{\sigma}_{\alpha}^{2}$ are given in (\ref{eq:muasiga}).
The update of the posterior mean $\boldsymbol{\mu}_{\alpha}=\mathbf{W}_{\alpha}^{-1}\boldsymbol{b}_{\alpha}$
involves a high-dimensional matrix inverse $\mathbf{W}_{\alpha}^{-1}$,
where $\mathbf{W}_{\alpha}\in\mathbb{C}^{K\left(Q+1\right)\times K\left(Q+1\right)}$.
In the following, we propose a subspace constrained matrix inverse
method together with a robust design to greatly reduce the complexity
for updating $\boldsymbol{\mu}_{\alpha}$. 

Let $\hat{\mathcal{S}}$ denote the estimated support of $\boldsymbol{\alpha}$
obtained in the previous iteration, as will be detailed soon. Define
a vector $\boldsymbol{\mu}_{\alpha}^{s}$ such that
\begin{equation}
\boldsymbol{\mu}_{\alpha,\hat{\mathcal{S}}}^{s}=\left\langle \gamma\right\rangle \mathbf{W}_{\alpha,\hat{\mathcal{S}}}^{-1}\left\langle \mathbf{A}_{\alpha,\hat{\mathcal{S}}}^{H}\right\rangle \boldsymbol{y},\label{eq:mu0}
\end{equation}
and $\boldsymbol{\mu}_{\alpha,\hat{\mathcal{S}}^{c}}^{s}=\boldsymbol{0}$,
where $\boldsymbol{\mu}_{\alpha,\hat{\mathcal{S}}}^{s}\in\mathbb{C}^{\left|\hat{\mathcal{S}}\right|\times1}$,
$\mathbf{W}_{\alpha,\hat{\mathcal{S}}}\in\mathbb{C}^{\left|\hat{\mathcal{S}}\right|\times\left|\hat{\mathcal{S}}\right|}$,
and $\mathbf{A}_{\alpha,\hat{\mathcal{S}}}\in\mathbb{C}^{MN\left(T_{p}+T_{d}\right)\times\left|\hat{\mathcal{S}}\right|}$
are subvector/submatrices of $\boldsymbol{\mu}_{\alpha}^{s}$, $\mathbf{W}_{\alpha}$,
and $\mathbf{A}_{\alpha}$ with the element/column/row indices lying
in $\hat{\mathcal{S}}$, and $\boldsymbol{\mu}_{\alpha,\hat{\mathcal{S}}^{c}}^{s}$
contains the rest elements of $\boldsymbol{\mu}_{\alpha}^{s}$. If
$\hat{\mathcal{S}}=\mathcal{S}$, $\boldsymbol{\mu}_{\alpha}^{s}$
is expected to provide a good initial solution for minimizing $\varphi_{\alpha}\left(\boldsymbol{u}_{\alpha}\right)$
in (\ref{eq:muasiga}) based on the gradient method. 

To improve the robustness of the algorithm against the support estimation
error, we choose the best initial point $\boldsymbol{\mu}_{\alpha}^{0}$
from the following two choices: 1) $\boldsymbol{\mu}_{\alpha}^{s}$
obtained by the subspace constrained matrix inverse as in (\ref{eq:mu0});
2) the posterior mean $\boldsymbol{\mu}_{\alpha}^{p}$ from the previous
iteration, depending on which choice gives a lower value of the objective
function $\varphi_{\alpha}\left(\boldsymbol{\mu}_{\alpha}^{0}\right)$.
In the first few iterations, $\boldsymbol{\mu}_{\alpha}^{s}$ is expected
to be a good choice because the subspace constrained matrix inverse
can be used to accelerate the initial convergence speed, compared
to the MM-based methods in \cite{Duan_IFSBL,Xu_Turbo-IFVBI,Xu_SLA_VBI}
which completely avoid any matrix inverse. As the iterations go on,
the posterior mean $\boldsymbol{\mu}_{\alpha}^{p}$ from the previous
iteration becomes more and more accurate and it may also become a
good initial point.

Then we apply the gradient update for $B_{\alpha}\geq1$ times. Specifically,
in the $i\textrm{-th}$ gradient iteration, the posterior mean is
updated as
\begin{align}
\boldsymbol{\mu}_{\alpha}^{\left(i\right)} & =\boldsymbol{\mu}_{\alpha}^{\left(i-1\right)}-\epsilon_{\alpha}^{\left(i\right)}\nabla_{\boldsymbol{u}}\varphi_{\alpha}\left(\boldsymbol{u}_{\alpha}\right)\mid_{\boldsymbol{u}_{\alpha}=\boldsymbol{\mu}_{\alpha}^{\left(i-1\right)}},\label{eq:Gradx}
\end{align}
where $\epsilon_{\alpha}^{\left(i\right)}$ is the step size determined
by the Armijo rule, 
\begin{equation}
\nabla_{\boldsymbol{u}}\varphi_{\alpha}\left(\boldsymbol{u}_{\alpha}\right)\mid_{\boldsymbol{u}_{\alpha}=\boldsymbol{\mu}_{\alpha}^{\left(i-1\right)}}=\mathbf{W}_{\alpha}\boldsymbol{\mu}_{\alpha}^{\left(i-1\right)}-\boldsymbol{b}_{\alpha},
\end{equation}
and $B_{\alpha}$ is chosen to achieve a good trade-off between the
per iteration complexity and convergence speed. Such a robust design
ensures that the objective value $\varphi_{\alpha}\left(\boldsymbol{u}_{\alpha}\right)$
is decreased monotonically in each iteration until convergence to
a stationary solution, as proved in Theorem \ref{thm:Convergence-of-SC-VBI}.

Once we obtain the estimated posterior mean $\boldsymbol{\mu}_{\alpha}$,
we can obtain an estimated support $\hat{\mathcal{S}}$ for the next
iteration by finding the elements in $\boldsymbol{\mu}_{\alpha}$
with sufficiently large energy. Specifically, we set a small threshold
$\varepsilon>0$ and compare each element of $\boldsymbol{\mu}_{\alpha}$
with the threshold to obtain the estimated support as
\begin{equation}
\hat{\mathcal{S}}\triangleq\left\{ k,q\mid\forall\left|\boldsymbol{\mu}_{\alpha,k,q}\right|^{2}>\varepsilon\right\} ,\label{eq:S_hat}
\end{equation}
where the threshold $\varepsilon$ is chosen according to the noise
power $1/\left\langle \gamma\right\rangle $. A good choice for $\varepsilon$
is 2 to 3 times the noise power $1/\left\langle \gamma\right\rangle $.

In the first iteration, we can simplify set $\hat{\mathcal{S}}$ according
to the available prior information in practice or a simple baseline
algorithm (e.g., using orthogonal matching pursuit (OMP) \cite{Tropp_CE_OMP}).
In Section \ref{sec:SMUSIC-SCVBI-Algorithm-for}, we will propose
a coarse estimation algorithm to further reduce the number of required
grid points $Q$. In this case, $\hat{\mathcal{S}}$ in the first
iteration can be obtained according to the coarse estimation result.

\paragraph{Update Equation for $q\left(\boldsymbol{\rho}\right)$}

$q\left(\boldsymbol{\rho}\right)$ can be derived as
\begin{equation}
q\left(\boldsymbol{\rho}\right)=\prod_{k=1}^{K}\prod_{q=1}^{Q}\Gamma\left(\rho_{k,q};\check{a}_{k,q},\check{b}_{k,q}\right),\label{eq:poster_rho}
\end{equation}
where the approximate posterior parameters are given by:
\begin{align}
\check{a}_{k,q}= & \left\langle s_{k,q}\right\rangle a_{k,q}+\left\langle 1-s_{k,q}\right\rangle \overline{a}_{k,q}+1,\label{eq:ab_tilde}\\
\check{b}_{k,q}= & \left\langle s_{k,q}\right\rangle b_{k,q}+\left\langle 1-s_{k,q}\right\rangle \overline{b}_{k,q}+\left\langle \left|\alpha_{k,q}\right|^{2}\right\rangle .\nonumber 
\end{align}

\paragraph{Update Equation for $q\left(\boldsymbol{s}\right)$}

$q\left(\boldsymbol{s}\right)$ can be derived as 
\begin{equation}
q\left(\boldsymbol{s}\right)=\prod_{k=1}^{K}\prod_{q=1}^{Q}\left(\check{\lambda}_{k,q}\right)^{s_{k,q}}\left(1-\check{\lambda}_{k,q}\right)^{1-s_{k,q}},\label{eq:q(s)}
\end{equation}
where $\check{\lambda}_{k,q}$ is given by
\begin{equation}
\check{\lambda}_{k,q}=\frac{\lambda_{k,q}C_{k,q}}{\lambda_{k,q}C_{k,q}+\left(1-\lambda_{k,q}\right)\overline{C}_{k,q}},\label{eq:lambda_n_tilde}
\end{equation}
with $\lambda_{k,q}=\textrm{h}_{k,q}^{\text{A}}\left(s_{k,q}=1\right)$,
$C_{k,q}=\dfrac{b_{k,q}^{a_{k,q}}}{\Gamma\left(a_{k,q}\right)}\exp\left(\left(a_{k,q}-1\right)\left\langle \ln\rho_{k,q}\right\rangle -b_{n}\left\langle \rho_{k,q}\right\rangle \right)$,
and $\overline{C}_{k,q}=\dfrac{\overline{b}_{k,q}^{\overline{a}_{k,q}}}{\Gamma\left(\overline{a}_{k,q}\right)}\exp\left(\left(\overline{a}_{k,q}-1\right)\left\langle \ln\rho_{k,q}\right\rangle -\overline{b}_{k,q}\left\langle \rho_{k,q}\right\rangle \right)$.
Here, $\Gamma\left(\cdot\right)$ denotes the gamma function.

\paragraph{Update Equation for $q\left(\boldsymbol{x}\right)$}

From Lemma \ref{lem:ICVBI}, $q\left(\boldsymbol{x}\right)=\prod_{t_{d}=1}^{T_{d}}\prod_{n=1}^{N}q\left(\boldsymbol{x}_{t_{d},n}\right)$
with
\begin{equation}
q\left(\boldsymbol{x}_{t_{d},n}\right)=\mathcal{CN}\left(\boldsymbol{x}_{t_{d},n};\boldsymbol{\mu}_{x,t_{d},n},\boldsymbol{\sigma}_{x,t_{d},n}^{2}\right),\label{eq:poster_x-1}
\end{equation}
where $\boldsymbol{\mu}_{x,t_{d},n}=\mathbf{W}_{x,t_{d},n}^{-1}\boldsymbol{b}_{x,t_{d},n}$
and $\boldsymbol{\sigma}_{x,t_{d},n}^{2}$ are given in (\ref{eq:muxsigx}). 

\paragraph{Update Equation for $q\left(\gamma\right)$}

The posterior distribution $q\left(\gamma\right)$ is given by
\begin{equation}
q\left(\gamma\right)=\Gamma\left(\gamma;\check{c},\check{d}\right),\label{eq:q(gamma)}
\end{equation}
where the parameters $\check{c}$ and $\check{d}$ are given by
\begin{align}
\check{c} & =c+\left(T_{p}+T_{d}\right)MN,\label{eq:gamma_post_old}\\
\check{d} & =d+\left\langle \left\Vert \boldsymbol{y}-\mathbf{A}_{\alpha}\boldsymbol{\alpha}\right\Vert ^{2}\right\rangle _{q\left(\boldsymbol{\alpha}\right)q\left(\boldsymbol{x}\right)}.\nonumber 
\end{align}

The expectations used in the above update equations are summarized
as follows:
\[
\left\langle \rho_{k,q}\right\rangle =\dfrac{\check{a}_{k,q}}{\check{b}_{k,q}},\text{ }\left\langle s_{k,q}\right\rangle =\check{\lambda}_{k,q},\text{ }\left\langle \gamma\right\rangle =\dfrac{\check{c}}{\check{d}},
\]
\[
\left\langle \alpha_{k,q}^{2}\right\rangle =\left|\mu_{\alpha,k,q}\right|^{2}+\sigma_{\alpha,k,q}^{2},\text{ }\left\langle \ln\rho_{k,q}\right\rangle =\psi\left(\check{a}_{k,q}\right)-\ln\check{b}_{k,q},
\]
where $\psi\left(\cdot\right)\triangleq d\ln\left(\Gamma\left(\cdot\right)\right)$
denotes the logarithmic derivative of the gamma function. Note that
the expectations $\left\langle \mathbf{A}_{\alpha}^{H}\mathbf{A}_{\alpha}\right\rangle $
and $\left\langle \mathbf{A}_{\alpha}\right\rangle $ are w.r.t. $q\left(\boldsymbol{x}\right)$,
and the expectations $\left\langle \mathbf{\Psi}_{n}^{H}\mathbf{\Psi}_{n}\right\rangle $
and $\left\langle \mathbf{\Psi}_{n}^{H}\right\rangle $ are w.r.t.
$q\left(\boldsymbol{\alpha}\right)$, and those expectations can be
easily calculated and are omitted for conciseness. 

\subsubsection{Convergence Analysis}

Despite the above approximations in the calculation of the posterior
mean $\boldsymbol{\mu}_{\alpha}=\mathbf{W}_{\alpha}^{-1}\boldsymbol{b}_{\alpha}$,
we can show that the proposed BiSVBI still converges to a stationary
point of $\mathcal{\mathscr{A}}_{\mathrm{VBI}}$, as stated in the
following theorem, whose proof can be found in Appendix B.
\begin{thm}
[Convergence of BiSVBI]\label{thm:Convergence-of-SC-VBI}The BiSVBI
algorithm monotonically decrease the KLD objective in $\mathcal{\mathscr{A}}_{\mathrm{VBI}}$,
and every limiting point $q^{*}\left(\boldsymbol{v}\right)=\prod_{k\in\mathcal{H}}q^{*}\left(\boldsymbol{v}^{k}\right)$
generated by the BiSVBI is a stationary solution of Problem $\mathcal{\mathscr{A}}_{\mathrm{VBI}}$.
\end{thm}

\subsection{Turbo-BiSVBI-M Step\label{subsec:Turbo-IF-VBI-M-Step}}

Since there is no closed-form expression of $\ln p\left(\boldsymbol{y},\boldsymbol{\xi}\right)$,
it is challenging to directly solve the maximization problem in (\ref{eq:M-step}).
To get around this problem, one common solution is to construct a
surrogate function of $\ln p\left(\boldsymbol{y},\boldsymbol{\xi}\right)$
and maximize the surrogate function with respect to $\boldsymbol{\xi}$.
Specifically, in the $t\textrm{-th}$ iteration, the surrogate function
inspired by the EM method is given by
\begin{align}
Q\left(\boldsymbol{\xi};\boldsymbol{\xi}^{\left(t\right)}\right)= & \int_{\boldsymbol{v}}q^{\left(t\right)}\left(\overline{\boldsymbol{v}}\right)\ln\frac{p\left(\boldsymbol{y},\overline{\boldsymbol{v}};\boldsymbol{\xi}\right)}{q^{\left(t\right)}\left(\overline{\boldsymbol{v}}\right)}+\ln p\left(\boldsymbol{\xi}\right)\nonumber \\
= & -\sum_{t_{p}=1}^{T_{p}}\left\langle \gamma\right\rangle \left\Vert \boldsymbol{y}_{t_{p}}^{p}-\mathbf{U}_{t_{p}}\boldsymbol{\Phi}\left(\boldsymbol{\xi}\right)\boldsymbol{\mu}_{\alpha}^{\left(t\right)}\right\Vert ^{2}\nonumber \\
 & -\sum_{t_{d}=1}^{T_{d}}\left\langle \gamma\right\rangle \left\Vert \boldsymbol{y}_{t_{d}}^{d}-\left\langle \mathbf{X}_{t_{d}}\right\rangle \boldsymbol{\Phi}\left(\boldsymbol{\xi}\right)\boldsymbol{\mu}_{\alpha}^{\left(t\right)}\right\Vert ^{2}\nonumber \\
 & +\ln p\left(\boldsymbol{\xi}\right)+C_{2}\left(\boldsymbol{\xi}\right)+C_{0},\label{eq:func_Q}
\end{align}
where $q^{\left(t\right)}\left(\overline{\boldsymbol{v}}\right)\triangleq q\left(\overline{\boldsymbol{v}}\mid\boldsymbol{y};\boldsymbol{\xi}^{\left(t\right)}\right)$,
$q^{\left(t\right)}\left(\boldsymbol{\alpha}\right)=\mathcal{CN}\left(\boldsymbol{\mathbf{\alpha}};\boldsymbol{\mu}_{\alpha}^{\left(t\right)},\left(\boldsymbol{\sigma}_{\alpha}^{\left(t\right)}\right)^{2}\right)$,
and $q^{\left(t\right)}\left(\boldsymbol{x}_{t_{d},n}\right)=\mathcal{CN}\left(\boldsymbol{x}_{t_{d},n};\boldsymbol{\mu}_{x,t_{d},n}^{\left(t\right)},\left(\boldsymbol{\sigma}_{x,t_{d},n}^{\left(t\right)}\right)^{2}\right)$
are the approximate posteriors calculated in the latest E-step, $\left\langle \mathbf{X}_{t_{d}}\right\rangle $
is obtained by replacing all $x_{t_{d},k,n}$'s in $\mathbf{X}_{t_{d}}$
with its posterior mean $\mu_{t_{d},k,n}^{\left(t\right)}$'s, $C_{2}\left(\boldsymbol{\xi}\right)$
is related to the posterior variance $\left(\boldsymbol{\sigma}_{\alpha}^{\left(t\right)}\right)^{2}$
and $\left(\boldsymbol{\sigma}_{x,t_{d},n}^{\left(t\right)}\right)^{2}$,
and $C_{0}$ is a constant. 

When $q^{\left(t\right)}\left(\overline{\boldsymbol{v}}\right)=p\left(\overline{\boldsymbol{v}}\mid\boldsymbol{y};\boldsymbol{\xi}^{\left(t\right)}\right)$,
it can be shown that the gradient of $Q\left(\boldsymbol{\xi};\boldsymbol{\xi}^{\left(t\right)}\right)$
is consistent with $\ln p\left(\boldsymbol{y},\boldsymbol{\xi}\right)$
\cite{LiuAn_CE_Turbo_VBI}, and thus we can apply the gradient ascent
method to update $\boldsymbol{\xi}$ as
\begin{align}
\boldsymbol{\xi}^{\left(t+1\right)} & =\boldsymbol{\xi}^{\left(t\right)}+\varepsilon^{\left(t\right)}\frac{\partial Q\left(\boldsymbol{\xi};\boldsymbol{\xi}^{\left(t\right)}\right)}{\partial\boldsymbol{\xi}},\label{eq:update_r}
\end{align}
where $\varepsilon^{\left(t\right)}$ is the step size determined
by the Armijo rule. Such a gradient update method ensures that $\left\{ \boldsymbol{\xi}^{\left(t\right)}\right\} $
converges to a stationary solution of the MAP problem in (\ref{eq:M-step})
\cite{LiuAn_CE_Turbo_VBI}. In practice, $q^{\left(t\right)}\left(\overline{\boldsymbol{v}}\right)$
is not exactly equal to the true posterior $p\left(\overline{\boldsymbol{v}}\mid\boldsymbol{y};\boldsymbol{\xi}^{\left(t\right)}\right)$.
Moreover, it is also a good practice to omit the variance term $\frac{\partial C_{2}\left(\boldsymbol{\xi}\right)}{\partial\boldsymbol{\xi}}$
in the gradient calculation to reduce the complexity. Since the variational
posterior $q^{\left(t\right)}\left(\overline{\boldsymbol{v}}\right)$
obtained by the Turbo-BiSVBI algorithm is a good approximation for
$p\left(\overline{\boldsymbol{v}}\mid\boldsymbol{y};\boldsymbol{\xi}^{\left(t\right)}\right)$
and the variance term $C_{2}\left(\boldsymbol{\xi}\right)$ is usually
small, the more practical update in (\ref{eq:update_r}) based on
the approximate gradient calculated using $q^{\left(t\right)}\left(\overline{\boldsymbol{v}}\right)$
and omitting $\frac{\partial C_{2}\left(\boldsymbol{\xi}\right)}{\partial\boldsymbol{\xi}}$
can still find a good solution of (\ref{eq:M-step}), as shown in
the simulations.

The overall EM-Turbo-BiSVBI algorithm is summarized in Algorithm \ref{EM-Turbo-BiSVBI}.

\begin{algorithm}[t]
\begin{singlespace}
{\small\caption{\label{EM-Turbo-BiSVBI}EM-Turbo-BiSVBI algorithm}
}{\small\par}

\textbf{Input:} $\boldsymbol{y}$, initial parameters $\boldsymbol{\xi}^{\left(1\right)}$,
inner iteration numbers $I$, outer iteration number $T$ .

\textbf{Output:} $\hat{\boldsymbol{\alpha}}$, $\hat{\boldsymbol{x}}$,
and $\hat{\boldsymbol{\xi}}$.

\begin{algorithmic}[1]

\FOR{${\color{blue}{\color{black}t=1,\cdots,T}}$}

\STATE \textbf{Turbo-BiSVBI-E Step:}

\STATE \textbf{\% Module A: BiSVBI}

\FOR{${\color{blue}{\color{black}i=1,\cdots,I}}$}

\STATE Update $q^{\left(t\right)}\left(\boldsymbol{\alpha}\right)$
in (\ref{eq:poster_x}), where the posterior mean $\boldsymbol{\mu}_{\alpha}$
is obtained by performing the gradient update (\ref{eq:Gradx}) for
$B_{x}$ times with the initial point in (\ref{eq:mu0}).

\STATE Update $q^{\left(t\right)}\left(\boldsymbol{\rho}\right)$
using (\ref{eq:poster_rho}).

\STATE Update $q^{\left(t\right)}\left(\boldsymbol{s}\right)$ using
(\ref{eq:q(s)}).

\STATE Update $q^{\left(t\right)}\left(\boldsymbol{x}\right)$ using
(\ref{eq:poster_x-1}).

\STATE Update $q^{\left(t\right)}\left(\gamma\right)$ using (\ref{eq:q(gamma)}).

\ENDFOR

\STATE Calculate the MMSE estimates $\boldsymbol{\alpha}^{\left(t\right)}$,
$\boldsymbol{x}^{\left(t\right)}$, and $\gamma^{\left(t\right)}$,
and then update the support set $\hat{\mathcal{S}}$ using (\ref{eq:S_hat}).

\STATE Send the extrinsic messages $\left\{ \upsilon_{\eta_{k,q}\rightarrow s_{k,q}}\right\} $
to Module B, using (\ref{eq:extricsic message}).

\STATE\textbf{ \% Module B: Message passing}

\STATE Perform message passing over $\mathcal{G}_{\mathrm{B}}$ and
send the extrinsic messages $\left\{ \upsilon_{\beta_{k,q}\rightarrow s_{k,q}}\right\} $
to Module A.

\STATE \textbf{Turbo-BiSVBI-M Step:}

\STATE Construct the surrogate function $Q\left(\boldsymbol{\xi};\boldsymbol{\xi}^{\left(t\right)}\right)$
in (\ref{eq:func_Q}).

\STATE Update $\boldsymbol{\xi}^{\left(t+1\right)}$ via gradient
ascent, using (\ref{eq:update_r}).

\ENDFOR

\STATE\textbf{ }Output $\hat{\boldsymbol{\alpha}}=\boldsymbol{\alpha}^{\left(T\right)},$$\hat{\boldsymbol{x}}=\boldsymbol{x}^{\left(T\right)}$,
and $\hat{\boldsymbol{\xi}}=\boldsymbol{\xi}^{\left(T+1\right)}$.

\end{algorithmic}
\end{singlespace}
\end{algorithm}

\subsection{Complexity Analysis for Turbo-BiSVBI}

The complexity of Turbo-BiSVBI is dominated by the update of $q\left(\boldsymbol{\alpha}\right)$
and $q\left(\boldsymbol{x}\right)$. In the update of $q\left(\boldsymbol{\alpha}\right)$
in BiSVBI, the matrix inverse is constrained in the subspace of the
estimated support, and thus the complexity order is only $\mathcal{O}\left(\left|\hat{\mathcal{S}}\right|^{3}\right)$.
The complexity order of the matrix-vector multiplications in the gradient
update (\ref{eq:Gradx}) is $\mathcal{O}\left(K\left(Q+1\right)MN\left(T_{p}+T_{d}\right)\right)$.
In the update of $q\left(\boldsymbol{x}\right)$ in BiSVBI, the complexity
order of the matrix inverse and matrix multiplications to calculate
the posterior mean $\boldsymbol{\mu}_{x,t_{d},n}=\mathbf{W}_{x,t_{d},n}^{-1}\boldsymbol{b}_{x,t_{d},n}$
is $\mathcal{O}\left(M\left|\mathcal{K}_{t_{d},n}^{d}\right|^{2}+\left|\mathcal{K}_{t_{d},n}^{d}\right|^{3}\right)$.
Therefore, the overall complexity order per iteration of Turbo-BiSVBI
is 
\begin{align}
\mathcal{O}\left(K\left(Q+1\right)MN\left(T_{p}+T_{d}\right)+\left|\hat{\mathcal{S}}\right|^{3}\right.\nonumber \\
\left.+\sum_{t_{d}=1}^{T_{d}}\sum_{n=1}^{N}\left(M\left|\mathcal{K}_{t_{d},n}^{d}\right|^{2}+\left|\mathcal{K}_{t_{d},n}^{d}\right|^{3}\right)\right).\label{eq:CO0}
\end{align}

In the original update of $q\left(\boldsymbol{\alpha}\right)$ in
BiSVBI, the high-dimensional matrix inverse $\boldsymbol{\mu}_{\alpha}=\mathbf{W}_{\alpha}^{-1}\boldsymbol{b}_{\alpha}$
has a complexity order of $\mathcal{O}\left(K\left(Q+1\right)\right)^{3}$,
which is unacceptable in practice. The subspace constrained matrix
inverse in BiSVBI significantly reduces the complexity order to only
$\mathcal{O}\left(\Bigl|\hat{\mathcal{S}}\Bigr|^{3}\right)$, where
$\left|\hat{\mathcal{S}}\right|\ll K\left(Q+1\right)$. In addition,
the number of data streams per subcarrier $\left|\mathcal{K}_{t_{d},n}^{d}\right|\leq K$
is usually small (e.g., typically less than 8 in current systems).
Therefore, the complexity of the matrix inverse $\mathbf{W}_{x,t_{d},n}^{-1}$
in the update of $q\left(\boldsymbol{x}\right)$ is often acceptable
since the dimension of $\mathbf{W}_{x,t_{d},n}$ is only $\left|\mathcal{K}_{t_{d},n}^{d}\right|$. 
\begin{rem}
[Methods to further reduce the complexity]In practice, the $K$
users are often divided into $G$ groups such that only the users
in the same group will be allocated with the same pilot/data subcarriers.
In this case, $\mathbf{W}_{\alpha,\hat{\mathcal{S}}}^{-1}$, $\mathbf{W}_{\alpha}$,
and $\mathbf{A}_{\alpha}$ are all block diagonal matrices, and the
complexity can be further reduced by exploiting the block diagonal
structures. In addition, when the number of data streams $\left|\mathcal{K}_{t_{d},n}^{d}\right|$
per subcarrier becomes large (say up to 100 streams in future systems),
we can still avoid the matrix inverse in $\boldsymbol{\mu}_{x,t_{d},n}=\mathbf{W}_{x,t_{d},n}^{-1}\boldsymbol{b}_{x,t_{d},n}$
by first applying an inverse-free data recovery method \cite{Duan_IFSBL,Xu_SLA_VBI}
to generate a good initial point $\boldsymbol{\mu}_{x,t_{d},n}^{s}$
and then apply the gradient updates for $B_{x}$ times, similar to
the gradient update for $\boldsymbol{\mu}_{\alpha}^{\left(i\right)}$
in (\ref{eq:Gradx}). With similar robust design as for $q\left(\boldsymbol{\alpha}\right)$,
the convergence of BiSVBI is still guaranteed. Finally, the number
of grid points $Q$ to cover the entire 3D area is usually huge, and
thus the complexity order of the gradient update of $q\left(\boldsymbol{\alpha}\right)$
is still high. In the following section, we propose an efficient coarse
estimation algorithm to identify the angles of scatterers, which narrows
down the search region and provides more prior information for the
EM-Turbo-BiSVBI to significantly reduce the complexity and accelerate
the convergence speed.
\end{rem}

\section{SMUSIC-SCVBI Algorithm for Coarse Estimation\label{sec:SMUSIC-SCVBI-Algorithm-for}}

In this section, we propose a coarse estimation algorithm based on
SMUSIC \cite{TST-MUSIC} and SCVBI \cite{Liu_SCVBI} (called SMUSIC-SCVBI
algorithm) to estimate the angles for each user. The SMUSIC-SCVBI
algorithm first adopts SMUSIC to estimate the angles, and then refines
the angles using SCVBI. In the following, we shall focus on presenting
the SMUSIC-SCVBI algorithm for a reference user $k$. Note that in
the coarse estimation stage, the data is not recovered, and only the
pilot subcarriers can be used to estimate the delays. Since the number
of pilot subcarriers is usually small in practice, we do not estimate
the delays in the coarse estimation stage. However, if the number
of pilot subcarriers is sufficient, it is also easy to incorporate
the delay estimation into the coarse estimation stage by replacing
SMUSIC with spatial-temporary MUSIC (ST-MUSIC) \cite{TST-MUSIC}. 

\subsection{SMUSIC to Estimate the Angles}

In SMUSIC, the spatial-domain covariance matrix for user $k$ is first
calculated as
\begin{align}
\mathbf{R}^{s} & =\frac{1}{N_{s}}\sum_{t_{p}=1}^{T_{p}}\sum_{n\in\mathcal{N}_{t_{p},k}^{p}}\boldsymbol{y}_{t_{p},n}^{p}\left(\boldsymbol{y}_{t_{p},n}^{p}\right)^{H}\nonumber \\
 & +\frac{1}{N_{s}}\sum_{t_{d}=1}^{T_{d}}\sum_{n\in\mathcal{N}_{t_{d},k}^{d}}\boldsymbol{y}_{t_{d},n}^{d}\left(\boldsymbol{y}_{t_{d},n}^{d}\right)^{H}\nonumber \\
 & =\mathbf{V}_{s}^{s}\boldsymbol{\Lambda}_{s}^{s}\left(\mathbf{V}_{s}^{s}\right)^{H}+\mathbf{V}_{n}^{s}\boldsymbol{\Lambda}_{n}^{s}\left(\mathbf{V}_{n}^{s}\right)^{H},
\end{align}
where $N_{s}=\sum_{t_{p}=1}^{T_{p}}\left|\mathcal{N}_{t_{p},k}^{p}\right|+\sum_{t_{d}=1}^{T_{d}}\left|\mathcal{N}_{t_{d},k}^{d}\right|$
is the total number of available spatial samples, the column vectors
of $\mathbf{V}_{s}^{s}$ are the eigenvectors that span the signal
subspace of $\mathbf{R}^{s}$, corresponding to the largest $L_{k}$
eigenvalues. The number of multi-paths $L_{k}$ can be estimated using
the minimum descriptive length (MDL) criterion \cite{MDL}. And the
column vectors of $\mathbf{V}_{n}^{s}$ are the eigenvectors that
span the noise subspace of $\mathbf{R}^{s}$. $\boldsymbol{\Lambda}_{s}^{s},\boldsymbol{\Lambda}_{n}^{s}$
are diagonal matrices consisting of the associated eigenvalues. 

Then, the angles can be estimated at which the following pseudo-spectrum
achieve maximum values \cite{MUSICbase}:
\begin{eqnarray}
\mathcal{P}^{s}(\theta,\phi) & = & \frac{1}{\boldsymbol{a}(\theta,\phi)^{H}\mathbf{V}_{n}^{s}\left(\mathbf{V}_{n}^{s}\right)^{H}\boldsymbol{a}(\theta,\phi)},\label{eq:spectrium_s}
\end{eqnarray}
and let $\left(\tilde{\boldsymbol{\theta}}_{k},\tilde{\boldsymbol{\phi}}_{k}\right)=\left[\left(\tilde{\theta}_{k,1},\tilde{\phi}_{k,1}\right),...,\left(\tilde{\theta}_{k,\tilde{L}_{k}},\tilde{\phi}_{k,\tilde{L}_{k}}\right)\right]^{T}$
denote the $\tilde{L}_{k}$ angles found by the SMUSIC, i.e., $\tilde{L}_{k}$
angles achieving the maximum values of $\mathcal{P}^{s}(\theta,\phi)$.

\subsection{SCVBI to Refine the Angles}

Let $\mathcal{U}_{k}$ denote the set of users that share some data
subcarriers with user $k$. Clearly, the received signals $\boldsymbol{y}_{t_{d},n}^{d}$'s
on the data subcarriers $n\in\mathcal{N}_{t_{p},k}^{p}$ may also
contain the data signals from the other users in $\mathcal{U}_{k}$.
Therefore, the angles obtained by the SMUSIC may contain all angles
of the users in $\mathcal{U}_{k}$. Since the pilot subcarriers in
$\mathcal{N}_{t_{p},k}^{p}$'s only contain pilot signals from user
$k$, we can exploit the received pilot signals $\boldsymbol{y}_{t_{p},n}^{p}$'s
to identify the angles of user $k$ and filter out the angles of the
other users in $\mathcal{U}_{k}$.

Specifically, define the angular-domain sparse basis $\tilde{\mathbf{A}}\in\mathbb{C}^{M\times\tilde{L}_{k}}$
based on the estimated angles $\left(\tilde{\boldsymbol{\theta}}_{k},\tilde{\boldsymbol{\phi}}_{k}\right)$
as
\begin{equation}
\tilde{\mathbf{A}}\triangleq\left[\boldsymbol{a}\left(\tilde{\theta}_{k,1},\tilde{\phi}_{k,1}\right),\ldots,\boldsymbol{a}\left(\tilde{\theta}_{k,\tilde{L}_{k}},\tilde{\phi}_{k,\tilde{L}_{k}}\right)\right].
\end{equation}
Then the sparse representation of the angular-domain channel vector
on the $n\textrm{-th}$ subcarrier is given by
\begin{equation}
\boldsymbol{h}_{k,n}=\tilde{\mathbf{A}}\tilde{\boldsymbol{\alpha}}_{k,n},\label{eq:LDhNL-1}
\end{equation}
where $\tilde{\boldsymbol{\alpha}}_{k,n}=\left[\tilde{\alpha}_{k,n,1},...,\tilde{\alpha}_{k,n,\tilde{L}_{k}}\right]^{T}$
is the angular-domain channel vector on subcarrier $n$. Then the
received pilot signal can be rewritten as
\begin{equation}
\boldsymbol{y}_{t_{p},n}^{p}=\tilde{\mathbf{A}}\tilde{\boldsymbol{\alpha}}_{k,n}+\boldsymbol{z}_{t_{p},n}^{p},\forall t_{p},\forall n\in\mathcal{N}_{t_{p},k}^{p}.\label{eq:Psigk}
\end{equation}

Note that different $\tilde{\boldsymbol{\alpha}}_{k,n}$'s share a
common support because the channels on different subcarriers share
the same set of angles. As such, a similar three-layer BGG prior is
introduced to capture such a common sparsity. In particular, a common
support vector $\tilde{\boldsymbol{s}}_{k}\triangleq\left[\tilde{s}_{k,1},\ldots,\tilde{s}_{k,\tilde{L}_{k}}\right]^{T}\in\left\{ 0,1\right\} ^{\tilde{L}_{k}}$
is introduced to indicate the non-zero elements in $\tilde{\boldsymbol{\alpha}}_{k,n}$'s.
Let $\tilde{\boldsymbol{\rho}}_{k,n}=\left[\tilde{\rho}_{k,n,1},...,\tilde{\rho}_{k,n,\tilde{L}_{k}}\right]^{T}$
denote the precision vector of $\tilde{\boldsymbol{\alpha}}_{k,n}$.
Then the joint distribution can be expressed as
\begin{align}
 & p\left(\left\{ \tilde{\boldsymbol{\alpha}}_{k,n}\right\} ,\left\{ \tilde{\boldsymbol{\rho}}_{k,n}\right\} ,\tilde{\boldsymbol{s}}_{k}\right)\label{eq:p(x,rou,s)}\\
= & p\left(\tilde{\boldsymbol{s}}_{k}\right)\prod_{n\in\mathcal{N}_{k}^{p}}p\left(\tilde{\boldsymbol{\rho}}_{k,n}\mid\tilde{\boldsymbol{s}}_{k}\right)p\left(\tilde{\boldsymbol{\alpha}}_{k,n}\mid\tilde{\boldsymbol{\rho}}_{k,n}\right),
\end{align}
where $\mathcal{N}_{k}^{p}=\cup_{t_{p}}\mathcal{N}_{t_{p},k}^{p}$,
$p\left(\tilde{\boldsymbol{s}}_{k}\right)=\prod_{l=1}^{\tilde{L}_{k}}\left(\tilde{\lambda}_{k,l}\right)^{\tilde{s}_{k,l}}\left(1-\tilde{\lambda}_{k,l}\right)^{1-\tilde{s}_{k,l}},$and
$\tilde{\lambda}_{k,l}$ is the probability of $\tilde{s}_{k,l}=1$,
which is set to be $\tilde{\lambda}_{k,l}=\frac{1}{\left|\mathcal{U}_{k}\right|}$
in the simulations. $p\left(\tilde{\boldsymbol{\rho}}_{k,n}\mid\tilde{\boldsymbol{s}}_{k}\right)$
is a conditional Gamma distribution such that the mean value of $\tilde{\rho}_{k,n,l}$
is set to be $\mathcal{O}\left(1\right)$ when $\tilde{s}_{k,l}=1$
and a large value when $\tilde{s}_{k,l}=0$. Finally, $p\left(\tilde{\alpha}_{k,n,l}\mid\tilde{\rho}_{k,n,l}\right)=\mathcal{CN}\left(\tilde{\alpha}_{k,n,l};0,\tilde{\rho}_{k,n,l}^{-1}\right),\forall l$.

With the linear observation model in (\ref{eq:Psigk}) and the BGG
sparse prior in (\ref{eq:p(x,rou,s)}), the problem of recovering
$\tilde{\boldsymbol{\alpha}}_{n,k}$'s becomes a structured compressive
sensing problem, which can be solved efficiently by the SCVBI algorithm
\cite{Liu_SCVBI}. SCVBI is a special case of BiSVBI and the details
are omitted for conciseness. Once we obtain the estimated posterior
mean $\tilde{\boldsymbol{\mu}}_{k,n}$ of $\tilde{\boldsymbol{\alpha}}_{k,n}$,
we can obtain an estimated support $\tilde{\mathcal{S}}_{k,n}$ by
finding the elements in $\boldsymbol{\mu}$ with sufficiently large
energy. Then the angles of user $k$ can be identified as $\left(\hat{\boldsymbol{\theta}}_{k},\hat{\boldsymbol{\phi}}_{k}\right)=\left[\left(\tilde{\theta}_{k,l},\tilde{\phi}_{k,l}\right)\right]_{l\in\tilde{\mathcal{S}}_{k,n}}^{T}$.

When the number of pilot subcarriers is small, the more precise angle
information provided by the SMUSIC based on all subcarriers including
the data subcarriers can significantly enhance the performance of
angle detection and estimation in the SCVBI step. In practice, we
can choose the total number of data subcarriers $N_{D}$ used in the
SMUSIC for all users to achieve a trade off between complexity and
performance. When the number of pilot subcarriers is sufficient, we
can even set $N_{D}=0$, in which case the SMUSIC step is completely
removed.

\subsection{Compact Construction of the Position Grid}

To reduce redundancy, we apply clustering to group similar angles
in $\left(\hat{\boldsymbol{\theta}},\hat{\boldsymbol{\phi}}\right)=\left\{ \left(\hat{\boldsymbol{\theta}}_{k},\hat{\boldsymbol{\phi}}_{k}\right),k=1,...,K\right\} $
and consolidate each angle cluster into a single representative angle
grid point, thereby avoiding dense angle grids to reduce the complexity
and also to avoid overfitting of the channel. Specifically, the clustering
of angles is based on an oversampled uniform angle grid as defined
in (\ref{eq:uniangrid}) and (\ref{eq:defg}), with $Q_{1}=N_{o}M_{x}$
and $Q_{2}=N_{o}M_{y}$, where $N_{o}>1$ can be chosen to achieve
a trade off between the grid resolution and complexity, and a typical
value of $N_{o}$ is 2 or 4. The angles that fall within the same
angle grid are considered to belong to the same cluster. 

Specifically, let $J$ denote the number of clusters, $\mathcal{A}_{j}^{c}$
denote the set of angles in the $j$-th cluster, $\left(\theta_{j,i}^{c},\phi_{j,i}^{c}\right),i\in\mathcal{A}_{j}^{c}$
denote the $i$-th angle in the $j$-th cluster and $\tilde{\alpha}_{j,i}^{c}$
denote the associated complex gain. Then the representative angle
$\left(\theta_{j}^{c},\phi_{j}^{c}\right)$ for cluster $j$ can be
obtained as
\begin{equation}
\left(\theta_{j}^{c},\phi_{j}^{c}\right)=\frac{\sum_{i\in\mathcal{A}_{j}^{c}}\left|\tilde{\alpha}_{j,i}^{c}\right|^{2}\left(\theta_{j,i}^{c},\phi_{j,i}^{c}\right)}{\sum_{i\in\mathcal{A}_{j}^{c}}\left|\tilde{\alpha}_{j,i}^{c}\right|^{2}}.\label{eq:vsm_weight_average}
\end{equation}

Finally, let $\mathcal{C}_{k}$ denote the set of clusters that contains
the angles of user $k$. Then the initial dynamic position grid for
user $k$ is given by the $Q_{k}=\left|\mathcal{C}_{k}\right|\times Q_{r}$
combinations between the $\left|\mathcal{C}_{k}\right|$ angle grid
points $\left\{ \left(\theta_{j}^{c},\phi_{j}^{c}\right),j\in\mathcal{C}_{k}\right\} $
and $Q_{r}$ uniform range grid points, and the union of the position
grid of all users is given by the $Q=J\times Q_{r}$ combinations
between the $J$ angle grid points $\left\{ \left(\theta_{j}^{c},\phi_{j}^{c}\right),j=1,...,J\right\} $
and $Q_{r}$ uniform range grid points.

\subsection{Complexity Analysis for the Overall Algorithm}

The complexity of SMUSIC is upper bounded by \cite{MUSICbase}
\begin{equation}
\mathcal{O}\left(KM^{3}+KQ_{s}M^{2}+N_{D}M^{2}\right),\label{eq:OC1}
\end{equation}
where $Q_{s}=\mathcal{O}\left(M\right)$ is the number of grid points
used to search the pseudo-spectrum in (\ref{eq:spectrium_s}). The
complexity of SCVBI is upper bounded by \cite{Liu_SCVBI}
\begin{equation}
\mathcal{O}\left(KMNT_{p}+M^{3}\right).\label{eq:OC2}
\end{equation}
With the compact position grid of $Q_{k}$ points for user $k$, the
complexity of the EM-Turbo-BiSVBI is 
\begin{align}
\mathcal{O}\left(\left(\sum_{k=1}^{K}Q_{k}+1\right)MN\left(T_{p}+T_{d}\right)+\left|\hat{\mathcal{S}}\right|^{3}\right.\nonumber \\
\left.+\sum_{t_{d}=1}^{T_{d}}\sum_{n=1}^{N}\left(M\left|\mathcal{K}_{t_{d},n}^{d}\right|^{2}+\left|\mathcal{K}_{t_{d},n}^{d}\right|^{3}\right)\right).\label{eq:OC3}
\end{align}
Therefore, the complexity of the overall algorithm is upper bounded
by the summation of (\ref{eq:OC1}) - (\ref{eq:OC3}), which is much
smaller than that of the original EM-Turbo-BiSVBI in (\ref{eq:CO0}).

\section{Simulations \label{sec:Applications}}

In this section, we demonstrate the advantages of the proposed algorithm
by comparison with the following baselines.
\begin{itemize}
\item \textbf{Variational Bayes\textquoteright{} joint channel estimation
and symbol decoding (VB-CESD) \cite{Thoota_JCDE_VBI1,Thoota_JCDE_VBI2}:}
It is the state-of-the-art method for joint angular-delay domain channel
estimation and data recovery.
\item \textbf{ST-MUSIC \cite{TST-MUSIC}: }Use ST-MUSIC to estimate the
angles and delays of channel paths, least square (LS) to estimate
the channel gain, and MMSE to estimate the data. 
\item \textbf{Two-stage framework \cite{Cui_JCDE_twostage}:} In stage 1,
use the SCVBI algorithm to estimate the location-domain channels;
in stage2, use MMSE to estimate the data.
\item \textbf{EM-Turbo-BiSVBI IID:} It is the proposed EM-Turbo-BiSVBI algorithm
with an i.i.d. prior (the factor graph has no joint support vector
$\boldsymbol{\overline{s}}$ and the elements of different $\boldsymbol{s}_{k}$'s
are i.i.d.).
\item \textbf{EM-Turbo-BiVBI:} This scheme is a minor variation of the proposed
scheme. The only difference is that it directly calculates the posterior
mean $\boldsymbol{\mu}_{\alpha}=\mathbf{W}_{\alpha}^{-1}\boldsymbol{b}_{\alpha}$
using the high-dimensional matrix inverse instead of the subspace
constrained matrix inverse.
\item \textbf{Genie-aided:} It is the proposed method when the user locations
and time offsets are perfectly known.
\end{itemize}

In the simulations, the BS is equipped with a UPA consisting of $M_{x}=8$
horizontal antennas and $M_{z}=8$ vertical antennas, for a total
of $M=64$ antennas. The number of OFDM subcarriers is $N=1024$,
the carrier frequency is $3.5\ \textrm{GHz}$, and the subcarrier
spacing is $f_{0}=30\text{ kHz}$. The users are divided into groups
of similar positions and each user group is assigned with the same
set of data subcarriers. In the simulations, we focus on a reference
user group containing $4$ users. There are $T_{p}=1$ pilot OFDM
symbol and $T_{d}=2$ data OFDM symbols. The $N^{p}$ pilot signals
for each user are generated with random phase under the unit power
constraint, and they are uniformly inserted in the $N$ subcarriers
without intersections. All users share the same $N^{d}$ data subcarriers
which are uniformly inserted in the $N$ subcarriers to enjoy the
frequency diversity gain. Moreover, the data signals for each user
are generated by a complex Gaussian distribution with zero mean and
unit variance. In the simulations, we fix $N^{d}=256$ and vary $N^{p}$
to show the effect of the number of pilot subcarriers on the performance.

The scatterers are distributed in a $100\text{ m}\times100\text{ m}\times10\text{ m}$
area. The BS is at coordinates $\left[0\text{ m},0\text{ m},0\text{ m}\right]^{T}$
and the mobile users are random distributed around coordinates $\left[100\text{ m},0\text{ m},0\text{ m}\right]^{T}$
with the $k$-th user position denoted as $\boldsymbol{p}_{u,k}=\left[p_{u,k}^{x},p_{u,k}^{y},0\right]^{T}$
in Cartesian coordinate system. We assume that the prior distribution
of $\boldsymbol{p}_{u}$ is $p^{x}\sim\mathcal{N}\left(p_{u,k}^{x},\sigma_{p}^{2}/2\right)$
and $p^{y}\sim\mathcal{N}\left(p_{u,k}^{y},\sigma_{p}^{2}/2\right)$,
where $\sigma_{p}^{2}$ is set as $1$. The channel is generated according
to the UMA scenario in the Quadriga model defined by 3GPP R16 specifications
\cite{3GPP_R16}.

The time offsets $\tau_{o,k}$'s are within $\left[\frac{-2}{B},\frac{2}{B}\right]$,
where $B=Nf_{0}$ denotes the total bandwidth. The number of azimuth
and elevation angle grid points is set to $Q_{1}=Q_{2}=16$ and the
number of range grid points is set to $Q_{r}=10$. We use the root
mean square error (RMSE) as the performance metric for scatterer localization
and the normalized mean square error (NMSE) as the performance metric
for channel estimation and data recovery.

\subsection{Convergence Behavior}

\begin{figure}
\begin{centering}
\includegraphics[width=70mm]{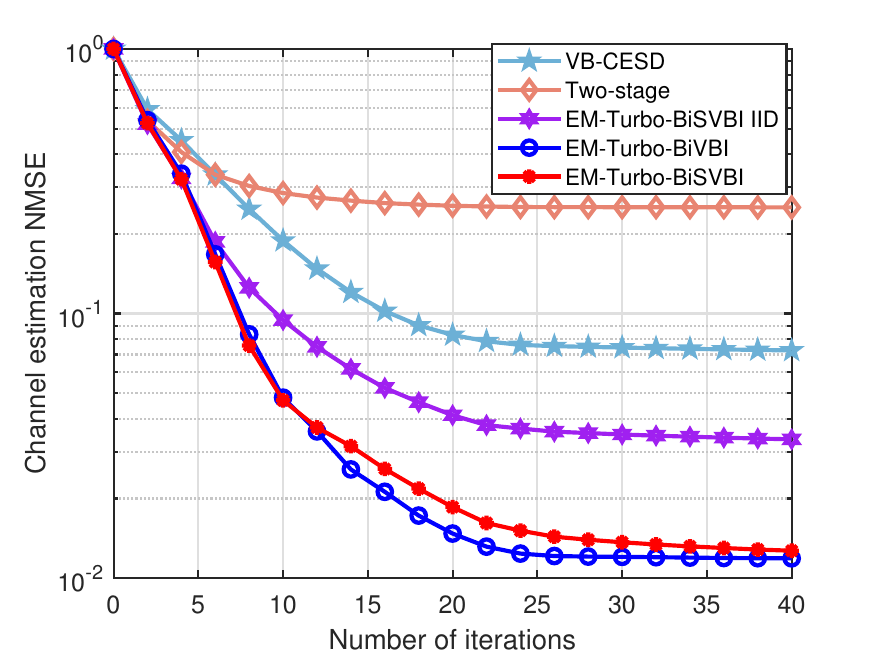}
\par\end{centering}
\caption{\label{fig:Convergence-Behavior:-channel}Convergence Behavior: Channel
estimation NMSE versus the EM iteration number.}

\begin{centering}
\includegraphics[width=70mm]{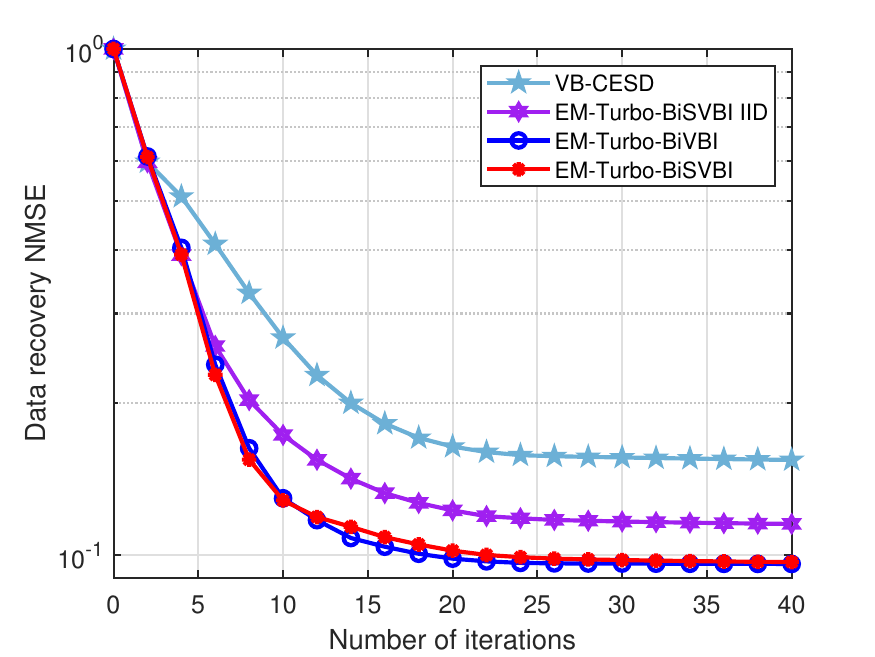}\caption{\label{fig:Convergence-Behavior:-Data}Convergence Behavior: Data
recovery NMSE versus the EM iteration number.}
\par\end{centering}
\end{figure}

In Fig. \ref{fig:Convergence-Behavior:-channel} and Fig. \ref{fig:Convergence-Behavior:-Data},
we compare the convergence behavior of different schemes when the
signal to noise ratio (SNR) is set to $-5\ \textrm{dB}$ and $N^{p}$
is set to $8$. The two-stage scheme converges very quickly but converges
to a relative poor stationary point. In contrast, other schemes based
on joint estimation converge within $25$ iterations and they find
better stationary solutions than the two-stage scheme. Moreover, both
the convergence speed and steady-state performance of the proposed
EM-Turbo-BiSVBI are similar to those of the more complicated EM-Turbo-BiVBI,
which means that the subspace constrained matrix inverse approach
can effectively reduce the computation overhead without performance
degradation.

\subsection{Impact of SNR}

\begin{figure*}[t]
\subfloat[\label{fig:CE_SNR}Channel estimation NMSE.]{\begin{centering}
\includegraphics[width=60mm]{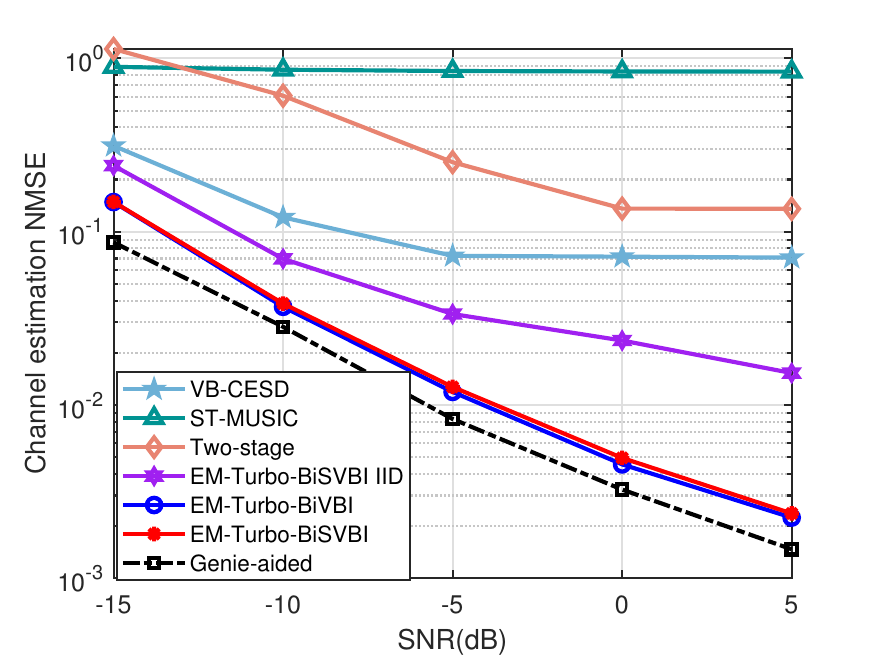}
\par\end{centering}
}\subfloat[\label{fig:DD_SNR}Data recovery NMSE.]{\begin{centering}
\includegraphics[width=60mm]{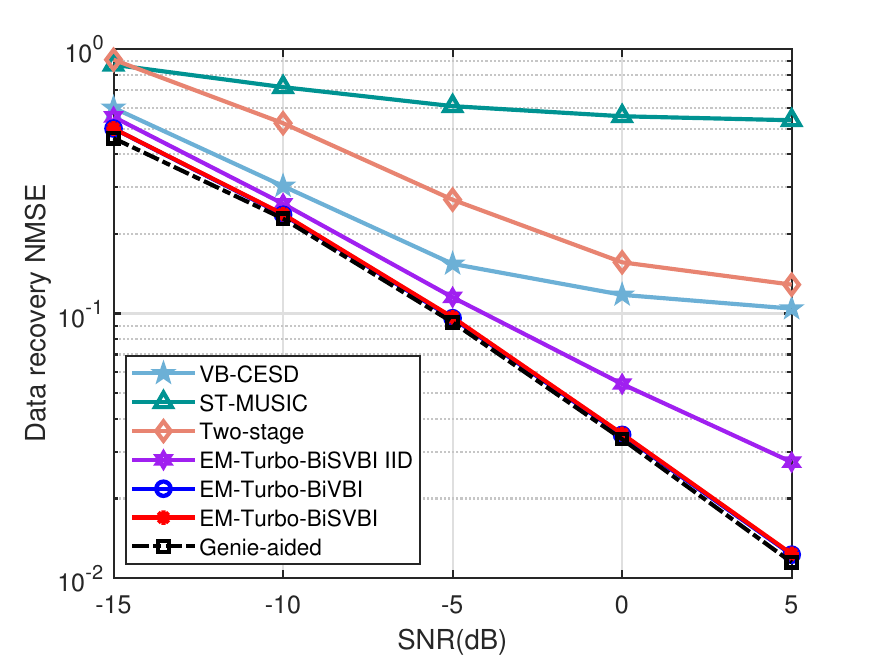}
\par\end{centering}
}\subfloat[\label{fig:RMSE_SNR}Scatterer localization RMSE.]{\begin{centering}
\includegraphics[clip,width=60mm]{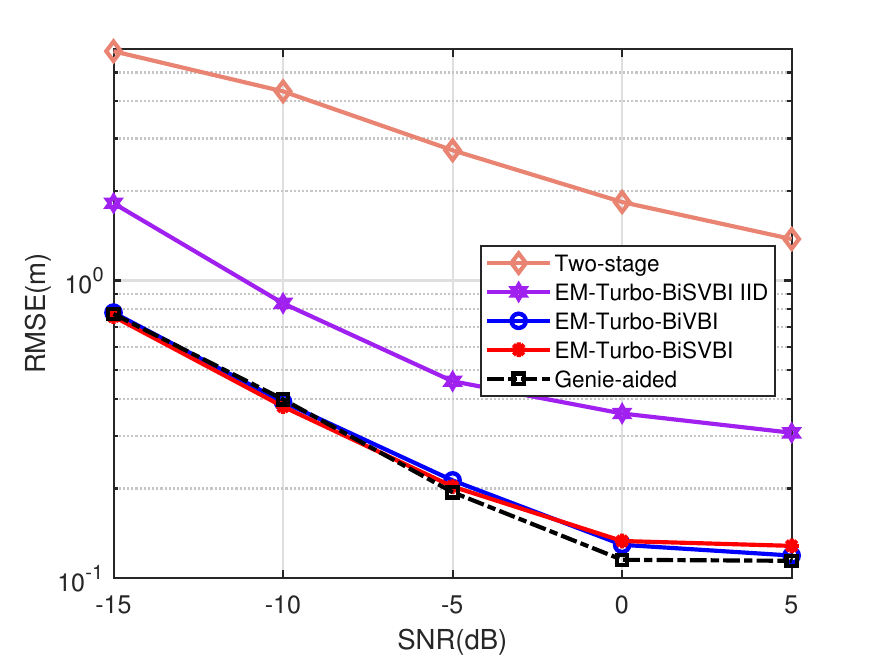}
\par\end{centering}
}\caption{Sensing and estimation performance versus SNR.}
\end{figure*}

In Fig. \ref{fig:CE_SNR}, Fig. \ref{fig:DD_SNR}, and Fig. \ref{fig:RMSE_SNR},
we set $N^{p}=8$ and evaluate the channel estimation NMSE, data recovery
NMSE, and scatterer localization RMSE performance under different
SNRs.\footnote{Since the VR-CESD and TS-MUSIC schemes estimate angles and delays
of channel paths directly, we do not plot the scatterer localization
RMSE performance for them.} It is obvious that the performance of all the schemes improves as
SNR increases. The ST-MUSIC scheme works very poorly since the estimation
error of delays is relatively large when the number of pilot subcarriers
is small. Besides, the joint estimation schemes achieve a significant
performance gain over the two-stage scheme. This is because channel
estimation and data recovery can be mutually enhanced based on both
received pilot and data signals. The EM-Turbo-BiSVBI IID scheme works
better than the VB-CESD scheme, which reflects the advantage of position-domain
channel modeling. Moreover, the performance gap between the EM-Turbo-BiSVBI
and EM-Turbo-BiSVBI IID is significant, which means that the proposed
three-layer BGG prior model can fully exploit the joint sparsity of
MU channels to enhance the sensing/estimation performance. Finally,
the performance of the proposed scheme is close to that of the genie-aided
scheme, especially in terms of data recovery NMSE and scatterer localization
RMSE, which implies that the non-ideal factors can be estimated accurate
enough by the Turbo-BiSVBI-M step.

\subsection{Impact of Number of Pilot Subcarriers}

\begin{figure*}[t]
\subfloat[\label{fig:CE_Np}Scatterer localization RMSE.]{\begin{centering}
\includegraphics[width=60mm]{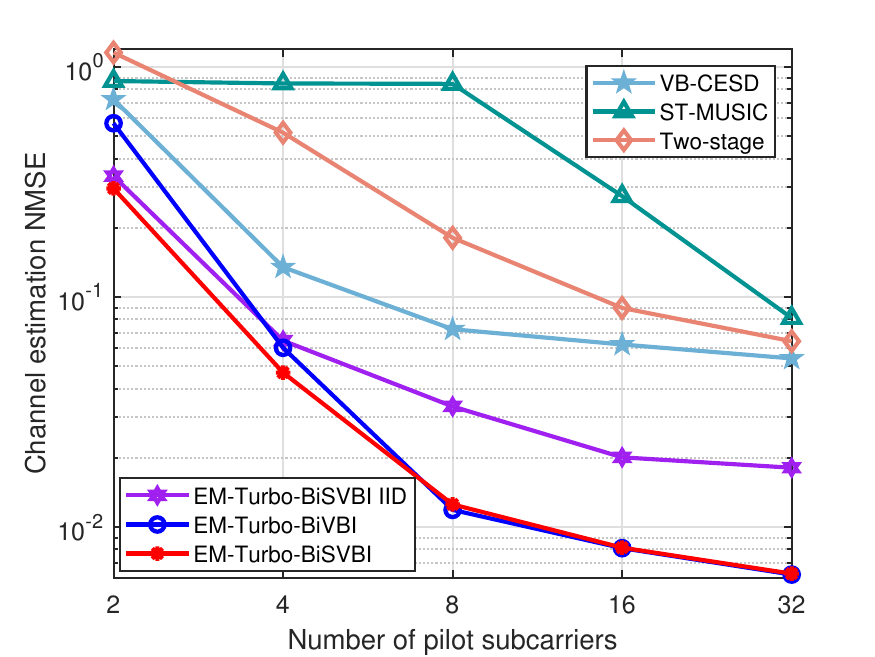}
\par\end{centering}
}\subfloat[\label{fig:DD_Np}Data recovery NMSE.]{\begin{centering}
\includegraphics[width=60mm]{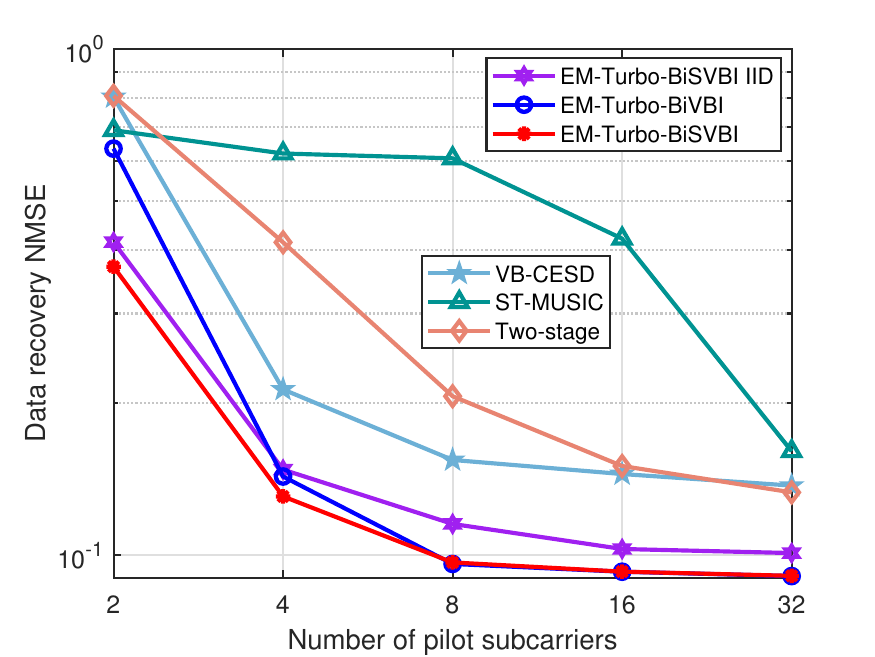}
\par\end{centering}
}\subfloat[\label{fig:RMSE_Np}Scatterer localization RMSE.]{\begin{centering}
\includegraphics[clip,width=60mm]{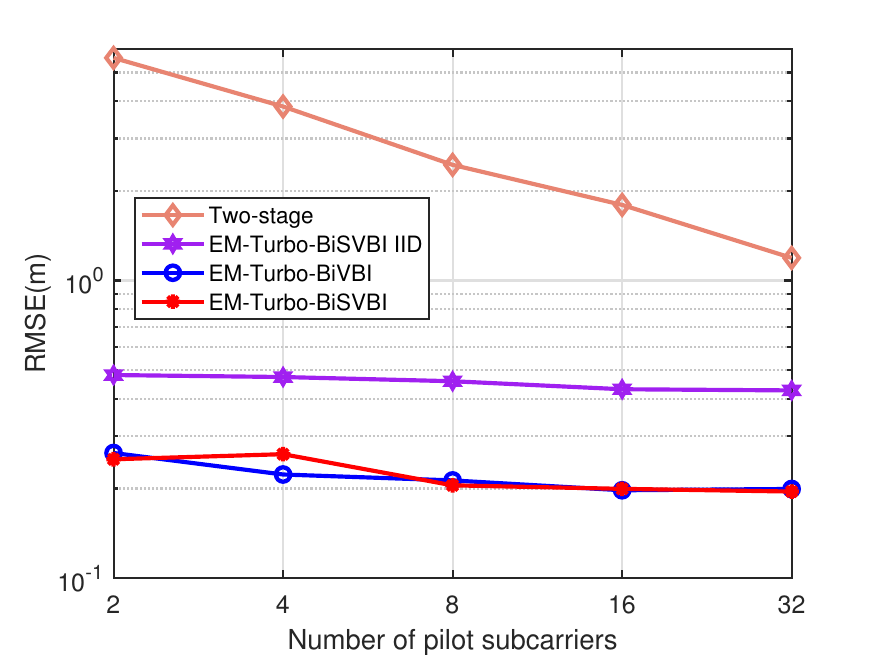}
\par\end{centering}
}\caption{Sensing and estimation performance versus the number of pilot subcarriers.}

\end{figure*}

In Fig. \ref{fig:CE_Np}, Fig. \ref{fig:DD_Np}, and Fig. \ref{fig:RMSE_Np},
we very $N^{p}$ from $2$ to $32$ to show how the number of pilot
subcarriers affects the sensing/estimation performance when $\textrm{SNR}=-5\ \textrm{dB}$.
The proposed EM-Turbo-BiSVBI achieves the best performance among all
the schemes. Besides, we find that the scatterer localization performance
of the proposed EM-Turbo-BiSVBI is not sensitive to the number of
pilot subcarriers. Even when $N^{p}=2$, its localization RMSE is
still very small. In other words, the proposed scheme can provide
high-accuracy localization service with very little pilot subcarriers
or even without any pilot.

\section{Conclusion\label{sec:Conlusion}}

This paper propose a novel joint sensing and data recovery algorithm
for massive MIMO-OFDM ISAC systems. A 3D location-domain sparse prior
model is firstly introduced to capture the joint sparsity of the MU
channel. Then, the joint problem is formulated as a bilinear structured
sparse recovery problem with a dynamic position grid and imperfect
parameters. An EM-Turbo-BiSVBI algorithm is proposed to solve this
joint sensing and data recovery problem, where the E-step performs
Bayesian estimation of the location-domain sparse MU channel by exploiting
the joint sparsity, and the M-step refines the dynamic position grid
and learns the imperfect factors via gradient update. In particular,
a BiSVBI algorithm is proposed to solve the bilinear sparse inference
problem in the E-step with both theoretical convergence guarantee
and low complexity. In addition, a SMUSIC-SCVBI algorithm is proposed
in the coarse estimation stage to narrow down the search range and
reduce the effective dimension of the location-domain channel, which
further reduces the complexity caused by the ultra-high-dimensional
sparse channel in the entire 3D space. The proposed algorithm is shown
by simulations to achieve much better trade-off between performance
and complexity than various baseline algorithms.

\appendix

\subsection{Proof of Lemma \ref{lem:ICVBI}\label{subsec:Proof-of-lemmaic}}

Given $q\left(\boldsymbol{x}\right),q\left(\boldsymbol{\rho}\right),q\left(\boldsymbol{s}\right),q\left(\gamma\right)$,
the KLD can be calculated as
\begin{align}
\textrm{KL}\left(q\parallel p\right)= & \int q\left(\boldsymbol{v}\right)\ln\frac{q\left(\boldsymbol{v}\right)}{\hat{p}\left(\boldsymbol{v}\mid\boldsymbol{y}\right)}\textrm{d}\boldsymbol{v}\nonumber \\
= & \left\langle \ln q\left(\boldsymbol{\alpha}\right)\right\rangle _{q\left(\boldsymbol{\alpha}\right)}-\left\langle \ln\hat{p}\left(\boldsymbol{v},\boldsymbol{y}\right)\right\rangle _{q\left(\boldsymbol{v}\right)}+C_{1}\nonumber \\
= & \left\langle \ln q\left(\boldsymbol{\alpha}\right)\right\rangle _{q\left(\boldsymbol{\alpha}\right)}-\left\langle \ln p\left(\boldsymbol{\alpha}\mid\boldsymbol{\rho}\right)\right\rangle _{q\left(\boldsymbol{\alpha}\right)q\left(\boldsymbol{\rho}\right)}\nonumber \\
 & -\left\langle \ln p\left(\boldsymbol{y}\mid\boldsymbol{\alpha},\boldsymbol{x},\gamma\right)\right\rangle _{q\left(\boldsymbol{\alpha}\right)q\left(\boldsymbol{x}\right)q\left(\gamma\right)}+C_{1},\nonumber \\
= & \left\langle \ln q\left(\boldsymbol{\alpha}\right)\right\rangle _{q\left(\boldsymbol{\alpha}\right)}+\left\langle \boldsymbol{\alpha}^{H}\textrm{Diag}\left(\left\langle \boldsymbol{\rho}\right\rangle \right)\boldsymbol{\alpha}\right\rangle _{q\left(\boldsymbol{\alpha}\right)}\nonumber \\
 & +\left\langle \gamma\right\rangle \left\langle \left\Vert \boldsymbol{y}-\mathbf{A}_{\alpha}\boldsymbol{\alpha}\right\Vert ^{2}\right\rangle _{q\left(\boldsymbol{\alpha}\right)q\left(\boldsymbol{x}\right)}+C_{1},\label{eq:KLD_q(alpha)}
\end{align}
where $C_{1}$ is a constant. Based on the constraint in (\ref{eq:GScon}),
Problem $\mathcal{\mathscr{A}}_{\mathrm{VBI}}$ is equivalent to finding
the optimal parameters of $q\left(\boldsymbol{\alpha}\right)$, denoted
by $\left\{ \ddot{\boldsymbol{\mu}}_{\alpha},\ddot{\boldsymbol{\sigma}}_{\alpha}^{2}\right\} $,
so that the KLD in (\ref{eq:KLD_q(alpha)}) is minimized. The equivalent
optimization problem is formulated as
\begin{align}
\ddot{\boldsymbol{\mu}}_{\alpha},\ddot{\boldsymbol{\sigma}}_{\alpha}^{2}= & \underset{\boldsymbol{u}_{\alpha},\boldsymbol{\sigma}_{\alpha}^{2}}{\arg\text{min}}\left[-\sum_{k=1}^{K}\sum_{q=0}^{Q}\ln\sigma_{\alpha,k,q}^{2}-\boldsymbol{\mu}_{\alpha}^{H}\textrm{Diag}\left(1/\boldsymbol{\sigma}_{\alpha}^{2}\right)\boldsymbol{\mu}_{\alpha}\right.\nonumber \\
 & +\left\langle \boldsymbol{\alpha}^{H}\left(\mathbf{W}_{\alpha}-\textrm{Diag}\left(1/\boldsymbol{\sigma}_{\alpha}^{2}\right)\right)\boldsymbol{\alpha}\right\rangle _{q\left(\boldsymbol{\alpha}\right)}\nonumber \\
 & \left.-\left\langle 2\mathfrak{Re}\left\{ \boldsymbol{\alpha}^{H}\left(\boldsymbol{b}_{\alpha}-\textrm{Diag}\left(1/\boldsymbol{\sigma}_{\alpha}^{2}\right)\boldsymbol{\mu}_{\alpha}\right)\right\} \right\rangle _{q\left(\boldsymbol{\alpha}\right)}\right]\nonumber \\
= & \underset{\boldsymbol{u}_{\alpha},\boldsymbol{\sigma}_{\alpha}^{2}}{\text{\ensuremath{\arg\text{min}}}}\left[-\sum_{k=1}^{K}\sum_{q=0}^{Q}\ln\sigma_{\alpha,k,q}^{2}-\boldsymbol{\mu}_{\alpha}^{H}\textrm{Diag}\left(1/\boldsymbol{\sigma}_{\alpha}^{2}\right)\boldsymbol{\mu}_{\alpha}\right.\nonumber \\
 & +\boldsymbol{\mu}_{\alpha}^{H}\left(\mathbf{W}_{\alpha}-\textrm{Diag}\left(1/\boldsymbol{\sigma}_{\alpha}^{2}\right)\right)\boldsymbol{\mu}_{\alpha}\nonumber \\
 & +\textrm{Tr}\left(\left(\mathbf{W}_{\alpha}-\textrm{Diag}\left(1/\boldsymbol{\sigma}_{\alpha}^{2}\right)\right)\textrm{Diag}\left(\boldsymbol{\sigma}_{\alpha}^{2}\right)\right)\nonumber \\
 & \left.-2\mathfrak{Re}\left\{ \boldsymbol{\mu}_{\alpha}^{H}\left(\boldsymbol{b}_{\alpha}-\textrm{Diag}\left(1/\boldsymbol{\sigma}_{\alpha}^{2}\right)\boldsymbol{\mu}_{\alpha}\right)\right\} \right]\nonumber \\
= & \underset{\boldsymbol{u}_{\alpha},\boldsymbol{\sigma}_{\alpha}^{2}}{\text{\ensuremath{\arg\text{min}}}}\left[\boldsymbol{\mu}_{\alpha}^{H}\mathbf{W}_{\alpha}\boldsymbol{\mu}_{\alpha}-2\mathfrak{Re}\left\{ \boldsymbol{\mu}_{\alpha}^{H}\boldsymbol{b}_{\alpha}\right\} \right.\nonumber \\
 & \left.+\sum_{k=1}^{K}\sum_{q=0}^{Q}\left(\mathrm{W}_{\alpha,k,q}\sigma_{\alpha,k,q}^{2}-\ln\sigma_{\alpha,k,q}^{2}\right)\right],
\end{align}
which is equal to (\ref{eq:muasiga}).

Similarly, given $q\left(\boldsymbol{\alpha}\right),q\left(\boldsymbol{\rho}\right),q\left(\boldsymbol{s}\right),q\left(\gamma\right)$,
the KLD can be calculated as
\begin{align}
\textrm{KL}\left(q\parallel p\right)= & \left\langle \ln q\left(\boldsymbol{x}\right)\right\rangle _{q\left(\boldsymbol{x}\right)}-\left\langle \ln p\left(\boldsymbol{x}\right)\right\rangle _{q\left(\boldsymbol{x}\right)}\nonumber \\
 & -\left\langle \ln p\left(\boldsymbol{y}\mid\boldsymbol{\alpha},\boldsymbol{x},\gamma\right)\right\rangle _{q\left(\boldsymbol{\alpha}\right)q\left(\boldsymbol{x}\right)q\left(\gamma\right)}+C_{2},\nonumber \\
= & \sum_{t=1}^{T_{d}}\sum_{n=1}^{N}\left[\left\langle \ln q\left(\boldsymbol{x}_{t_{d},n}\right)\right\rangle _{q\left(\boldsymbol{x}_{t_{d},n}\right)}\right.\nonumber \\
 & +\left\langle \boldsymbol{x}_{t_{d},n}^{H}\textrm{Diag}\left(\boldsymbol{\sigma}_{t_{d},n}^{2}\right)\boldsymbol{x}_{t_{d},n}\right\rangle _{q\left(\boldsymbol{x}_{t_{d},n}\right)}\nonumber \\
 & \left.+\left\langle \gamma\right\rangle \left\langle \left\Vert \boldsymbol{y}_{t_{d},n}^{d}-\mathbf{\Psi}_{n}\boldsymbol{x}_{t_{d},n}\right\Vert ^{2}\right\rangle _{q\left(\boldsymbol{\alpha}\right)q\left(\boldsymbol{x}_{t_{d},n}\right)}\right]+C_{2},\label{eq:KLD_q(x)}
\end{align}
where $C_{2}$ is a constant. Based on the constraint in (\ref{eq:GScon_x}),
Problem $\mathcal{\mathscr{A}}_{\mathrm{VBI}}$ is equivalent to finding
the optimal parameters of $q\left(\boldsymbol{x}\right)$, denoted
by $\left\{ \ddot{\boldsymbol{\mu}}_{x},\ddot{\boldsymbol{\sigma}}_{x}^{2}\right\} $,
so that the KLD in (\ref{eq:KLD_q(alpha)}) is minimized. The equivalent
optimization problem is formulated as
\begin{align}
 & \ddot{\boldsymbol{\mu}}_{x,t_{d},n},\ddot{\boldsymbol{\sigma}}_{x,t_{d},n}^{2}\nonumber \\
= & \underset{\boldsymbol{\mu}_{x,t_{d},n},\boldsymbol{\sigma}_{x,t_{d},n}^{2}}{\text{\ensuremath{\arg\text{min}}}}\left[\boldsymbol{\mu}_{x,t_{d},n}^{H}\mathbf{W}_{x,t_{d},n}\boldsymbol{\mu}_{x,t_{d},n}-2\mathfrak{Re}\left\{ \boldsymbol{\mu}_{x,t_{d},n}^{H}\boldsymbol{b}_{x,t_{d},n}\right\} \right.\nonumber \\
 & \left.+\sum_{k\in\mathcal{K}_{t_{d},n}^{d}}\left(\mathbf{W}_{x,t_{d},k,n}\boldsymbol{\sigma}_{x,t_{d},k,n}^{2}-\ln\boldsymbol{\sigma}_{x,t_{d},k,n}^{2}\right)\right],
\end{align}
which is equal to (\ref{eq:muxsigx}).

\subsection{Proof of Theorem \ref{thm:Convergence-of-SC-VBI} \label{subsec:Proof-of-TheoremSCVBI}}

The BiSVBI algorithm can be viewed as an alternating optimization
method to solve Problem $\mathcal{\mathscr{A}}_{\mathrm{VBI}}$. It
is clear that the BiSVBI can monotonically decreasing the KLD objective,
thus the KLD will converge to a limit. For convenience, we use 
\[
\boldsymbol{\xi}\triangleq\left\{ \boldsymbol{\mu}_{\alpha},\boldsymbol{\sigma}_{\alpha}^{2},\boldsymbol{\mu}_{x},\boldsymbol{\sigma}_{x}^{2},\check{a}_{k,q},\check{b}_{k,q},\check{\lambda}_{k,q},\check{c},\check{d}\mid\forall k,q\right\} 
\]
 to represent the parameters of $q\left(\boldsymbol{v}\right)$. Let
$\boldsymbol{\xi}_{j}$ denote the $j\textrm{-th}$ block of $\boldsymbol{\xi}$
for $j=1,\ldots,B$, where $B=\left|\boldsymbol{\xi}\right|$ . Then,
the KLD in (\ref{eq:KLDmin}) can be rewritten as a function of $\boldsymbol{\xi}$,
i.e.
\begin{equation}
\textrm{KL}\left(\boldsymbol{\xi}\right)=\int q\left(\boldsymbol{v};\boldsymbol{\xi}\right)\ln\frac{q\left(\boldsymbol{v};\boldsymbol{\xi}\right)}{\hat{p}\left(\boldsymbol{v}\mid\boldsymbol{y};\boldsymbol{\xi}\right)}\textrm{d}\boldsymbol{v}.
\end{equation}
In the following, we will prove that every limiting point $q^{*}\left(\boldsymbol{v};\boldsymbol{\xi}^{*}\right)$
generated by the BiSVBI is a stationary solution of Problem $\mathcal{\mathscr{A}}_{\mathrm{VBI}}$.

In the $t\textrm{-th}$ iteration of the BiSVBI, we update $\boldsymbol{\xi}_{j},\forall j$
alternatively as
\begin{align}
\boldsymbol{\xi}_{j}^{\left(t\right)} & =\begin{cases}
\underset{\boldsymbol{\xi}_{j}}{\textrm{argmin}}\textrm{KL}\left(\boldsymbol{\xi}_{j},\boldsymbol{\xi}_{-j}^{\left(t-1\right)}\right), & \textrm{if}\ \boldsymbol{\xi}_{j}\neq\boldsymbol{\mu}_{\alpha},\\
\boldsymbol{\mu}_{\alpha}^{\left(t\right)\left(B_{\alpha}\right)}, & \textrm{if}\ \boldsymbol{\xi}_{j}=\boldsymbol{\mu}_{\alpha},
\end{cases}\label{eq:update_rule}
\end{align}
where $\left(\cdotp\right)^{\left(t\right)}$ stands for the $t\textrm{-th}$
iteration, $\boldsymbol{\xi}_{-j}^{\left(t-1\right)}=\left(\boldsymbol{\xi}_{1}^{\left(t\right)},\ldots,\boldsymbol{\xi}_{j-1}^{\left(t\right)},\boldsymbol{\xi}_{j+1}^{\left(t-1\right)},\ldots,\boldsymbol{\xi}_{B}^{\left(t-1\right)}\right)$,
and $\boldsymbol{\mu}_{\alpha}^{\left(t\right)\left(B_{\alpha}\right)}$
is computed by the following gradient update for $B_{\alpha}$ times:
\begin{align}
\boldsymbol{\mu}_{\alpha}^{\left(t\right)\left(i\right)} & =\boldsymbol{\mu}_{\alpha}^{\left(t\right)\left(i-1\right)}-\epsilon_{\alpha}^{\left(t\right)\left(i\right)}\nabla_{\boldsymbol{\mu}_{\alpha}}\textrm{KL}\left(\boldsymbol{\mu}_{\alpha},\boldsymbol{\xi}_{-j}^{\left(t-1\right)}\right)\mid_{\boldsymbol{\mu}_{\alpha}=\boldsymbol{\mu}_{\alpha}^{\left(t\right)\left(i-1\right)}}\nonumber \\
 & =\boldsymbol{\mu}_{\alpha}^{\left(t\right)\left(i-1\right)}-\epsilon_{\alpha}^{\left(t\right)\left(i\right)}\nabla_{\boldsymbol{\mu}_{\alpha}}\varphi_{\alpha}\left(\boldsymbol{\mu}_{\alpha}\right)\mid_{\boldsymbol{\mu}_{\alpha}=\boldsymbol{\mu}_{\alpha}^{\left(t\right)\left(i-1\right)}},\label{eq:mu(t,i)}
\end{align}
for $i=1,\ldots,B_{\alpha}$, where $\epsilon_{\alpha}^{\left(t\right)\left(i\right)}$
is the step size determined by the Armijo rule. Based on the robust
design, the initial value $\boldsymbol{\mu}_{\alpha}^{\left(t\right)\left(0\right)}$
is given by
\begin{equation}
\boldsymbol{\mu}_{\alpha}^{\left(t\right)\left(0\right)}=\begin{cases}
\boldsymbol{\mu}_{\alpha}^{\left(t-1\right)}, & \textrm{if}\ \varphi\left(\boldsymbol{\mu}_{\alpha,\hat{\mathcal{S}}}^{s}\right)\geq\varphi\left(\boldsymbol{\mu}_{\alpha}^{\left(t-1\right)}\right),\\
\boldsymbol{\mu}_{\alpha,\hat{\mathcal{S}}}^{s}, & \textrm{otherwise}.
\end{cases}\label{eq:robust_design}
\end{equation}
where $\boldsymbol{\mu}_{\alpha,\hat{\mathcal{S}}}^{s}$ is obtained
by the subspace constrained matrix inverse in (\ref{eq:mu0}). The
robust design in (\ref{eq:robust_design}) can ensure $\textrm{KL}\left(\boldsymbol{\mu}_{\alpha}^{\left(t-1\right)},\boldsymbol{\xi}_{-j}^{\left(t-1\right)}\right)\geq\textrm{KL}\left(\boldsymbol{\mu}_{\alpha}^{\left(t\right)\left(0\right)},\boldsymbol{\xi}_{-j}^{\left(t-1\right)}\right)$,
where the equality holds when $\varphi\left(\boldsymbol{\mu}_{\alpha,\hat{\mathcal{S}}}^{s}\right)\geq\varphi\left(\boldsymbol{\mu}_{\alpha}^{\left(t-1\right)}\right)$.
Besides, the gradient update in (\ref{eq:mu(t,i)}) can ensure $\textrm{KL}\left(\boldsymbol{\mu}_{\alpha}^{\left(t\right)\left(i-1\right)},\boldsymbol{\xi}_{-j}^{\left(t-1\right)}\right)\geq\textrm{KL}\left(\boldsymbol{\mu}_{\alpha}^{\left(t\right)\left(i\right)},\boldsymbol{\xi}_{-j}^{\left(t-1\right)}\right),\forall i$,
where the equality holds only when the gradient w.r.t. $\boldsymbol{\mu}_{\alpha}$
is zero. Based on these, we have
\begin{align}
 & \textrm{KL}\left(\boldsymbol{\mu}_{\alpha}^{\left(t-1\right)},\boldsymbol{\xi}_{-j}^{\left(t-1\right)}\right)\geq\textrm{KL}\left(\boldsymbol{\mu}_{\alpha}^{\left(t\right)\left(0\right)},\boldsymbol{\xi}_{-j}^{\left(t-1\right)}\right)\geq\nonumber \\
 & \textrm{KL}\left(\boldsymbol{\mu}_{\alpha}^{\left(t\right)\left(1\right)},\boldsymbol{\xi}_{-j}^{\left(i-1\right)}\right)\geq\ldots\geq\textrm{KL}\left(\boldsymbol{\mu}_{\alpha}^{\left(t\right)\left(B_{\alpha}\right)},\boldsymbol{\xi}_{-j}^{\left(i-1\right)}\right)\nonumber \\
= & \textrm{KL}\left(\boldsymbol{\mu}_{\alpha}^{\left(t\right)},\boldsymbol{\xi}_{-j}^{\left(i-1\right)}\right).\label{eq:KL_inequality}
\end{align}
Now we explain the non-increasing property of the KLD in the $t\textrm{-th}$
iteration. For the case of $\boldsymbol{\xi}_{j}\neq\boldsymbol{\mu}_{\alpha}$,
it is clear that $\textrm{KL}\left(\boldsymbol{\xi}_{j},\boldsymbol{\xi}_{-j}^{\left(t-1\right)}\right)$
is minimized w.r.t. $\boldsymbol{\xi}_{j}$. While for the case of
$\boldsymbol{\xi}_{j}=\boldsymbol{\mu}_{\alpha}$, we have $\textrm{KL}\left(\boldsymbol{\mu}_{\alpha}^{\left(t-1\right)},\boldsymbol{\xi}_{-j}^{\left(t-1\right)}\right)\geq\textrm{KL}\left(\boldsymbol{\mu}_{\alpha}^{\left(t\right)},\boldsymbol{\xi}_{-j}^{\left(t-1\right)}\right)$,
i.e., the KLD is non-increasing. Therefore, the KLD will keep decreasing
until converging to a certain value, and we must have 
\begin{equation}
\lim_{t\rightarrow\infty}\nabla_{\boldsymbol{\xi}_{j}}\textrm{KL}\left(\boldsymbol{\xi}_{j},\boldsymbol{\xi}_{-j}^{\left(t-1\right)}\right)=0,\forall j.\label{eq:gradzero}
\end{equation}
 Otherwise, the KLD will keep decreasing to negative infinity, which
contradicts with the fact that $\textrm{KL}\left(\boldsymbol{\xi}\right)\geq0$.
Then according to (\ref{eq:gradzero}) and the property of gradient
update, we must have $\lim_{t\rightarrow\infty}\left\Vert \boldsymbol{\mu}_{\alpha}^{\left(t\right)}-\boldsymbol{\mu}_{\alpha}^{\left(t-1\right)}\right\Vert =0$.
Moreover, it follows from (\ref{eq:gradzero}) and the strong convexity
of $\textrm{KL}\left(\boldsymbol{\xi}_{j},\boldsymbol{\xi}_{-j}^{\left(t-1\right)}\right)$
w.r.t. $\boldsymbol{\xi}_{j},\forall\boldsymbol{\xi}_{j}\neq\boldsymbol{\mu}_{\alpha}$
that $\lim_{t\rightarrow\infty}\left\Vert \boldsymbol{\xi}_{j}^{\left(t\right)}-\boldsymbol{\xi}_{j}^{\left(t-1\right)}\right\Vert =0,\forall\boldsymbol{\xi}_{j}\neq\boldsymbol{\mu}_{\alpha}$.
Therefore, we have
\begin{equation}
\lim_{t\rightarrow\infty}\left\Vert \boldsymbol{\xi}_{j}^{\left(t\right)}-\boldsymbol{\xi}_{j}^{\left(t-1\right)}\right\Vert =0,\forall j.\label{eq:continuexi}
\end{equation}
It follows from (\ref{eq:continuexi}) that all the $B$ sequences
$\left\{ \boldsymbol{\xi}_{j}^{\left(t\right)},\boldsymbol{\xi}_{-j}^{\left(t\right)}\right\} ,j=1,...,B$
have the same set of limiting points. Let $\left\{ \boldsymbol{\xi}_{j}^{\left(t_{k}\right)},\boldsymbol{\xi}_{-j}^{\left(t_{k}\right)},k=1,2,...\right\} $
denote a subsequence that converges to a limiting point $\boldsymbol{\xi}^{*}$.
Suppose $\boldsymbol{\xi}^{*}$ is not a stationary point of $\textrm{KL}\left(\boldsymbol{\xi}\right)$,
then $\nabla_{\boldsymbol{\xi}}\textrm{KL}\left(\boldsymbol{\xi}\right)\neq0$
and it follows from (\ref{eq:continuexi}) that $\lim_{k\rightarrow\infty}\nabla_{\boldsymbol{\xi}_{j}}\textrm{KL}\left(\boldsymbol{\xi}_{j},\boldsymbol{\xi}_{-j}^{\left(t_{k}\right)}\right)\neq0$
must hold at least for some $j$, which contradicts with (\ref{eq:gradzero}).
Therefore, every limiting point $\boldsymbol{\xi}^{*}$ must be a
stationary point of $\textrm{KL}\left(\boldsymbol{\xi}\right)$. In
other word, every limiting point $q^{*}\left(\boldsymbol{v};\boldsymbol{\xi}^{*}\right)$
generated by the BiSVBI is a stationary solution of Problem $\mathcal{\mathscr{A}}_{\mathrm{VBI}}$.

\bibliographystyle{IEEEtran}
\bibliography{ISAC,Others,Localization_CE,CS_methods}

\end{document}